\documentclass[a4paper,USenglish,cleveref,autoref,thm-restate]{lipics-v2021}

\nolinenumbers

\usepackage{subcaption}
\usepackage{mathtools}

\definecolor{darkcyan}{rgb}{0.0, 0.55, 0.55}

\definecolor{darkred}{RGB}{200,0,0}

\title{Computing Non-Obtuse Triangulations\\ with Few Steiner Points} 

\titlerunning{Computing Non-Obtuse Triangulations with Few Steiner Points} 
\author{Mikkel Abrahamsen}{University of Copenhagen, Denmark}{miab@di.ku.dk}{https://orcid.org/0000-0003-2734-4690}{Supported by Independent Research Fund Denmark, grant 1054-00032B, and by the Carlsberg Foundation, grant CF24-1929.}
\author{Florestan Brunck}{University of Copenhagen, Denmark}{flbr@di.ku.dk}{https://orcid.org/0000-0003-4921-2824}{Supported by a DNRF Chair grant from the Danish National Research Foundation and partially by the BARC grant of the Villum Foundation.
}
\author{Jacobus Conradi}{University of Bonn, Germany}{conradi@cs.uni-bonn.de}{https://orcid.org/0000-0002-8259-1187}{Funded by the iBehave Network: Sponsored by the Ministry of Culture and Science of the State of North Rhine-Westphalia.}
\author{Benedikt Kolbe}{University of Bonn, Hausdorff Center for Mathematics, Lamarr Institute for Machine Learning and Artificial Intelligence, Germany}{bkolbe@uni-bonn.de}{https://orcid.org/0009-0005-0440-4912}{This research has partly been funded by the Federal Ministry of Education and Research of Germany
and the state of North-Rhine Westphalia as part of the Lamarr-Institute for Machine Learning and
Artificial Intelligence.} 
\author{André Nusser}{Université Côte d'Azur, CNRS, Inria, France}{andre.nusser@cnrs.fr}{https://orcid.org/0000-0002-6349-869X}{Supported by the France 2030 investment plan managed by the ANR as part of the Initiative of Excellence of Université Côte d'Azur with reference number ANR-15-IDEX-01.}

\authorrunning{M. Abrahamsen, F. Brunck, J. Conradi, B. Kolbe and A. Nusser} 

\Copyright{M. Abrahamsen, F. Brunck, J. Conradi, B. Kolbe and A. Nusser} 



\category{CG Challenge} 

\relatedversion{} 

\supplement{}
\supplementdetails[subcategory={Source Code}, cite={}, swhid={}]{Software}{https://github.com/JacobusTheSecond/nonObtuseTri}

\acknowledgements{We thank the Challenge organizers and other competitors for their
time, feedback, and making this whole event possible. We thank Max Kanold for invaluable input enabling high quality visualizations which made later discussions on how to improve our algorithm all the more fruitful. We gratefully acknowledge the access to the Marvin cluster of the University of Bonn. }

\nolinenumbers 

\EventEditors{John Q. Open and Joan R. Access}
\EventNoEds{2}
\EventLongTitle{42nd Conference on Very Important Topics (CVIT 2016)}
\EventShortTitle{CVIT 2016}
\EventAcronym{CVIT}
\EventYear{2016}
\EventDate{December 24--27, 2016}
\EventLocation{Little Whinging, United Kingdom}
\EventLogo{}
\SeriesVolume{42}
\ArticleNo{23}

\begin{document}

\maketitle

\begin{abstract}
We present the winning implementation of the Seventh Computational Geometry Challenge\break\hbox{(CG:SHOP 2025)}.
The task in this challenge was to find non-obtuse triangulations for given planar regions, respecting a given set of constraints consisting of extra vertices and edges that must be part of the triangulation.
The goal was to minimize the number of introduced Steiner points.
Our approach is to maintain a constrained Delaunay triangulation, for which we repeatedly remove, relocate, or add Steiner points.
We use local search to choose the action that improves the triangulation the most, until the resulting triangulation is non-obtuse.
\end{abstract}

\begin{figure}[!bh]
    \centering
    \includegraphics[width=\linewidth]{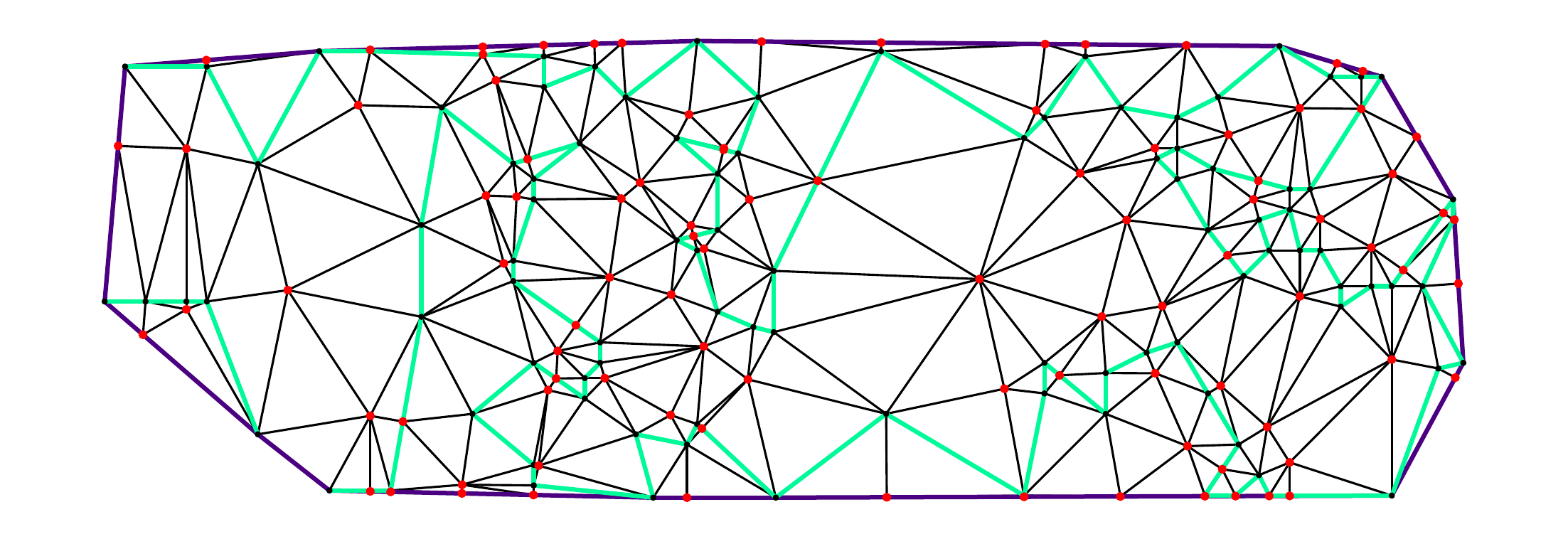}
    \caption{A non-obtuse triangulation produced by our implementation.
    Here, as in the rest of the paper, the region to be triangulated is enclosed by the fat curve, the given vertices are black, and the constraint edges are green.
    The Steiner points are red.}
    \label{fig:firstex}
\end{figure}
\newpage
\section{Introduction}

The Computational Geometry Challenge (CG:SHOP) is an annual competition on geometric optimization. The 2025 edition involved finding non-obtuse triangulations of a given planar straight-line graph (PSLG), a planar embedding of a graph where edges are represented as line segments; see \Cref{fig:firstex}. 
The input consists of a PSLG $G=(V,E)$ and a simple polygon $P$ with vertices in $V$.
The goal is to output a non-obtuse triangulation $\mathcal T$ of $P$ with vertices $V'$ so that $V\subseteq V'$ and the edges of $\mathcal T$ cover $E$.
The vertices $V' \setminus V$ are called \emph{Steiner points} and the objective is to minimize them.
We refer to the survey~\cite{survey} for further details.

The problem of finding non-obtuse triangulations has a long history, dating back to the 1960s~\cite{BZ,circlepacking, grid, maehara, saraf,higher-dim}. In the 2-dimensional setting, the latest development is Bishop's algorithm --- the only one to provably output a non-obtuse triangulation of any PSLG with a number of Steiner points polynomial in the vertices of the input. In general, the computational complexity of minimizing the number of Steiner points or triangles remains unknown, and the problem may be intractable already for simple polygons. 

In this paper, we present our winning implementation. The instance set of the challenge consisted of $150$ instances with $|V|$ and $|E|$ as large as $250$. Our computed solutions were the best for 116 instances, while the runner-up team, `Gwamegi'~\cite{gwamegi25}, was best in 4. For the remaining 30 instances, we matched Gwamegi in number of Steiner points.

\section{Algorithm}

Let us first outline the idea behind our algorithm.
Since a non-obtuse triangulation is a Delaunay triangulation (the local Delaunay condition is respected by the inscribed angle theorem), our algorithm maintains a (constrained) Delaunay triangulation (CDT) that gets modified successively.
Whenever a obtuse triangle is present, the algorithm makes a modification by removing, relocating, or adding a Steiner point and updating the triangulation accordingly.
More precisely, the algorithm operates in rounds, and in each round we consider a set of candidates for local changes and choose the most promising one.
In each round, the input is the original instance and the current set of Steiner points, which have been added in the previous rounds.
We compute a set of candidate actions, of which there are three types: (i) the insertion of a new Steiner point, (ii) the relocation of an existing Steiner point, and (iii) the replacement of two adjacent Steiner points with a single Steiner point.
For each action, we compute the resulting triangulation after applying the action.
Each of the resulting triangulations is evaluated using a cost function (given in \cref{sec:evaluation}, Equation~\eqref{eq:evalfunction}), which involves the number of Steiner points and the number of obtuse triangles of different types.
The output produced by the round is the triangulation that minimizes the cost.
The process terminates when the CDT has no obtuse triangles left.
In particular, our algorithm produces a non-obtuse triangulation whenever it terminates.

The way in which we insert a new Steiner point is inspired by the work of Erten and Ungör~\cite{ungörerten1,ungörerten2, erten-not}.
They described a general approach to finding triangulations avoiding small or large angles, where new Steiner points are chosen as optimal points along the Voronoi diagram (see \cite[Section 2.1]{erten-not}). We extend this idea to a visibility-constrained Voronoi diagram~\cite{pslgBoundedVoronoiDiagram} which takes into account the constraints of the PSLG.

\subsection{Primitives}\label{sec:primitives}

\begin{figure}[tb]
    \centering
    \includegraphics[width=0.8\linewidth]{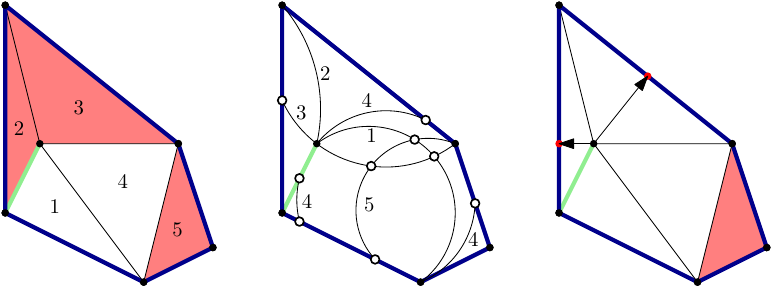}
    \caption{\emph{left:} CDT with obtuse triangles marked in red; \emph{middle:} its \texttt{circleArrangement}; \emph{right:} the two \texttt{altitudeDrop} points of the two upper obtuse triangles and its resulting CDT.}
    \label{fig:arrangementAndAltitude}
\end{figure}

\begin{figure}[tb]
    \centering
    \includegraphics[width=\linewidth]{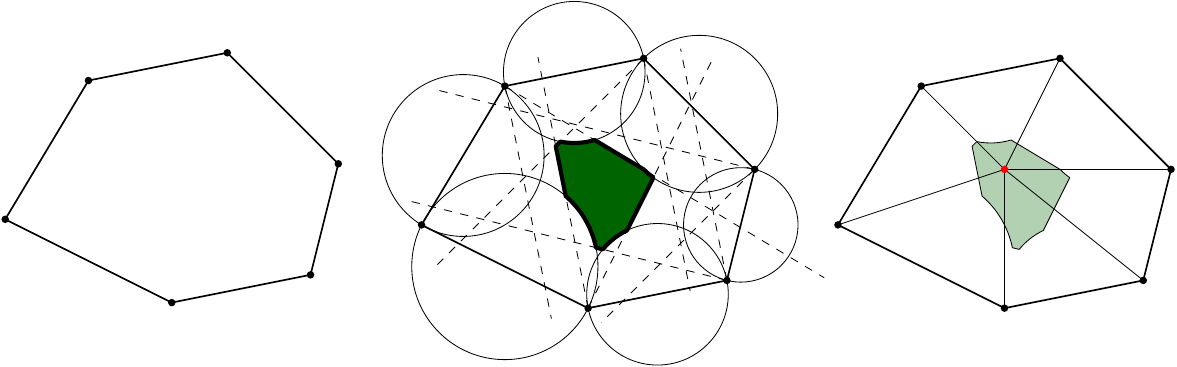}
    \caption{Illustration of the computation of the \texttt{polygonCenter} primitive.}
    \label{fig:polygon-solver}
\end{figure}

\begin{figure}[tb]
    \centering
    \includegraphics[width=0.8\linewidth]{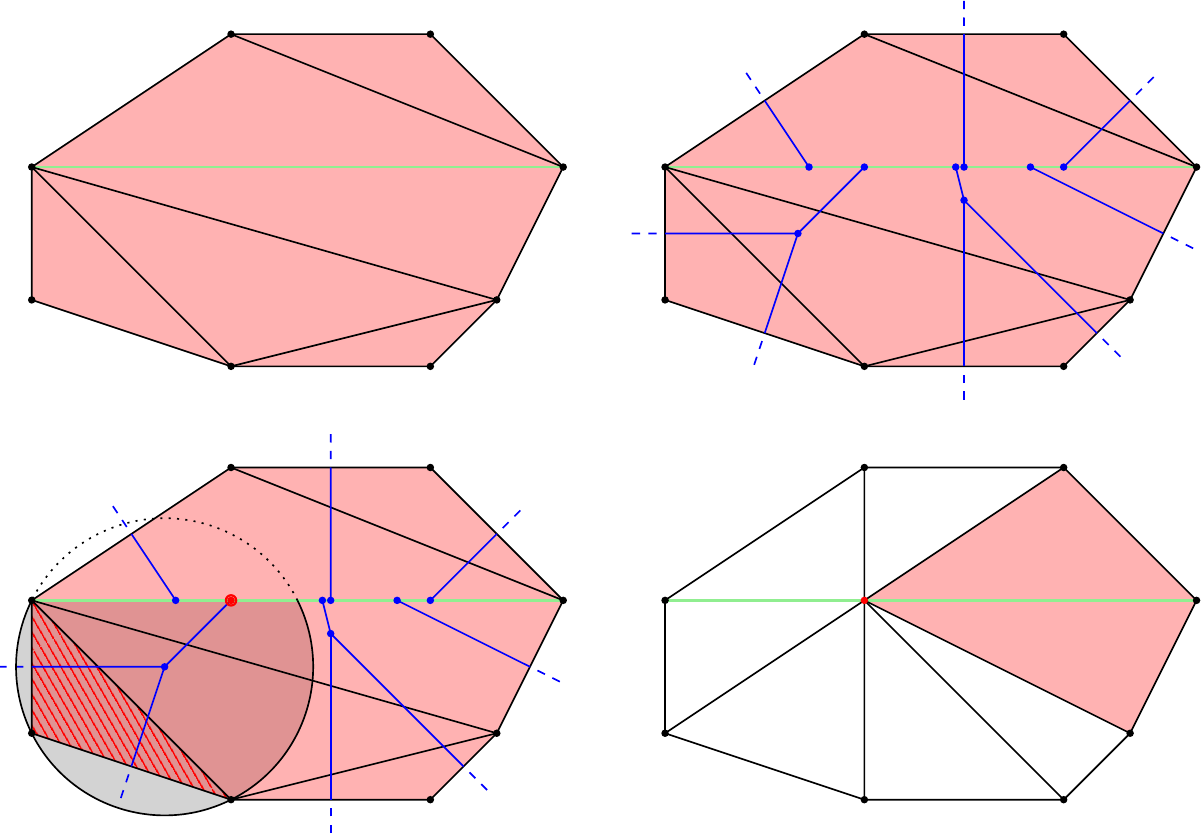}
    \caption{Center identification in the style of Erten and Ungör~\cite{ungörerten1,ungörerten2, erten-not}, from the visibility-constrained Voronoi diagram \cite{pslgBoundedVoronoiDiagram}.}
    \label{fig:center-identification}
\end{figure}

We first describe the geometric primitives that we are basing our algorithm on:

\begin{itemize}
    \item \texttt{altitudeDrop}: Given an obtuse triangle, compute the unique point on the longest side that splits the obtuse triangle into two right-angled triangles (\cref{fig:arrangementAndAltitude}, right).
    \item \texttt{polygonCenter}: Given a simple polygon, compute a center point such that connecting each vertex of the polygon with the center point results in a non-obtuse triangulation of the polygon, or decide no such point exists. Such a point can be computed as follows: Observe that for any two points $p$ and $q$, a point $x$ forms a non-obtuse triangle $pqx$ if and only if $x$ lies inside the slab defined by the two hyperplanes rooted at $p$ and $q$ with normal $q-p$ and outside the disk centered at $\frac{p+q}{2}$ with radius $\frac{\|q-p\|}{2}$. Pick any point in the intersection of these sets defined by all consecutive pairs of vertices $p$ and $q$ on the boundary of the polygon (\cref{fig:polygon-solver}).
    \item \texttt{visibilityVoronoi}: Given a CDT, compute the visibility-constrained Voronoi diagram (sometimes called bounded Voronoi diagram) \cite{pslgBoundedVoronoiDiagram}. (\cref{fig:center-identification}, top right).
    \item \texttt{clippedCircumcircle}: Let $T$ be a triangle in a CDT and $C$ its circumcircle. The disk bounded by $C$ may be partitioned into multiple pieces by the constraints of the CDT. Compute the piece of the partition containing $T$ (\cref{fig:center-identification}, bottom left).
    \item \texttt{circleArrangement}: Given a CDT, compute first the \texttt{clippedCircumcircle} for each triangle, then the arrangement of these clipped circumcircles (\cref{fig:arrangementAndAltitude}, middle).
\end{itemize}

\subsection{Action Generation}\label{sec:actions}

In this section we build on the geometric primitives we just introduced, and describe in more detail the three types of candidate actions that the algorithm evaluates for a provided CDT $\mathcal{D}$, which together form the action set $\mathcal{A}(\mathcal{D})$ of $\mathcal{D}$. Every Steiner point, triangle, and edge of $\mathcal{D}$ and every cell in the \texttt{circleArrangement} of $\mathcal{D}$ has at most one action associated to it. Every action includes two steps: the insertion, relocation or deletion of Steiner points and ensuring that the maintained triangulation is a CDT. We now describe the types of actions.

\subparagraph{Insertion} Let $T$ be a triangle in $\mathcal{D}$. If $T$ is non-obtuse already, then it has no action associated to it. Hence, let $T$ be obtuse instead. If the longest side of $T$ (opposite the unique obtuse angle in $T$) is a constrained edge of $\mathcal{D}$, the candidate action associated to $T$ adds the unique point described by the \texttt{altitudeDrop} primitive.
If instead the longest side of $T$ is not constrained, the candidate action associated to $T$ adds the point $v$ in the \texttt{clippedCircumcircle} of $T$ that lies on an edge $e$ of the \texttt{visibilityVoronoi} of $\mathcal D$ and maximizes the distance to the points of $\mathcal{D}$ defining $e$ (see \cref{{fig:center-identification}}). 
This point $v$ can efficiently be identified via a local search in the \texttt{visibilityVoronoi} diagram. We observe that $v$ maximizes the distance to all other points in the CDT 
while still lying in the \texttt{clippedCircumcircle} of $T$. This guarantees that its insertion into the CDT results in the removal
 of $T$ from the CDT, accompanied by an intuitively increased likelihood that only few angles of triangles incident to $v$ are obtuse (c.f.~\cite{ungörerten1,ungörerten2, erten-not}).

\begin{figure}
    \centering
    \includegraphics[width=\linewidth]{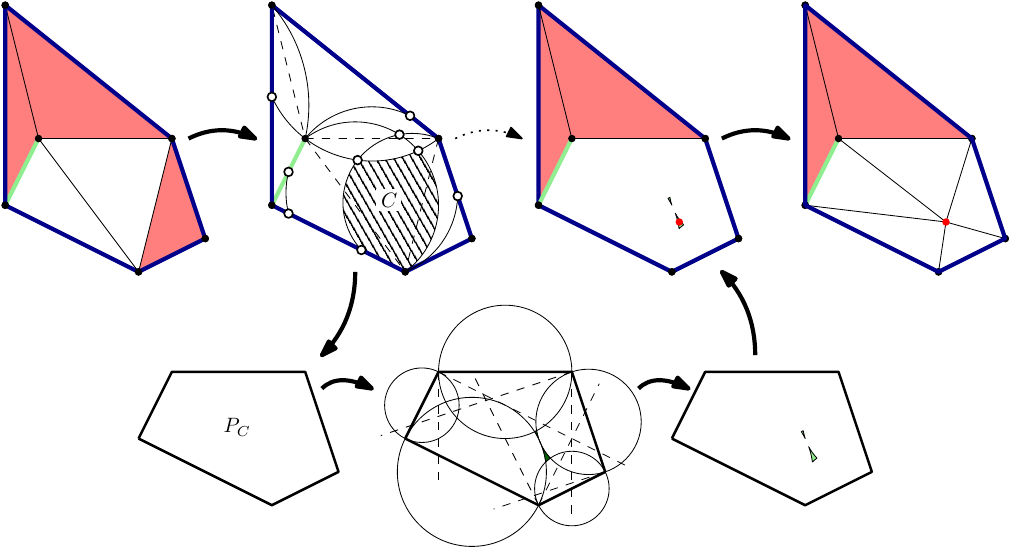}
    \caption{Center identification by choosing a cell $C$ from the \texttt{circleArrangement} and computing the \texttt{polygonCenter} of its corresponding polygon $P_C$.
    }
    \label{fig:circleArrangementCenter}
\end{figure}

Next, let $C$ be a cell in the \texttt{circleArrangement}.
Let $\{T_1,\ldots\}$ be the set of all triangles whose \texttt{clippedCircumcircle} contains $C$. By classical Delaunay arguments, the union of triangles in $\{T_1,\ldots\}$ defines a polygon $P_C$. 
We note that the CDT resulting from adding a point $v\in C$ is obtained by replacing the edges of $\mathcal D$ in $P_C$ with the segments $vp$ for every vertex $p$ of $P_C$.
In particular, if at least one triangle in $\{T_1,\ldots\}$ is obtuse and the \texttt{polygonCenter} $v$ of $P_C$ lies in $C$, then adding $v$ to the CDT ensures that the number of obtuse triangles in the CDT strictly decreases.
Thus, if at least one $T_i$ is obtuse and the \texttt{polygonCenter} of $P_C$ exists,
the candidate action associated to $C$ adds this \texttt{polygonCenter} (see \cref{fig:circleArrangementCenter}), otherwise no action is associated to $C$.

\subparagraph{Relocation} Let $v$ be a Steiner point of $\mathcal{D}$. If $v$ is not part of an obtuse triangle, $v$ has no action associated to it.
Hence, let $v$ be part of some obtuse triangle instead.
Let $P$ be the polygon resulting from the union of all triangles that $v$ is a part of.
Assume the \texttt{polygonCenter} of $P$ exists, otherwise $v$ has no action associated to it.
The action associated to $v$ moves $v$ to the \texttt{polygonCenter} of $P$.

\subparagraph{Deletion} Let $e$ be an edge in $\mathcal{D}$ defined by two Steiner points $v$ and $w$.
The union of all triangles that $v$ or $w$ are a part of forms a polygon $P$.
Assume the \texttt{polygonCenter} of $P$ exists, otherwise $e$ has no action associated to it. The action associated to $e$ removes the two points $v$ and $w$ from $\mathcal{D}$ and adds the \texttt{polygonCenter} of $P$.

\subsection{Evaluation}\label{sec:evaluation}

The evaluation function of a CDT $\mathcal{D}$ consists of two parts: A naive evaluation function and an algorithm that improves the accuracy of the naive evaluation. 

\begin{figure}[!htp]
    \centering
    
    \subfloat[\nolinenumbers\centering Initial CDT with hatched non-superseded triangles and how triangles are superseded. \\$\mathrm{eval}(\mathcal{D})=19.9$, $\mathrm{eval}_3(\mathcal{D})=14.5$.\label{fig4:0}]{\scalebox{0.45}{\includegraphics[width=\textwidth]{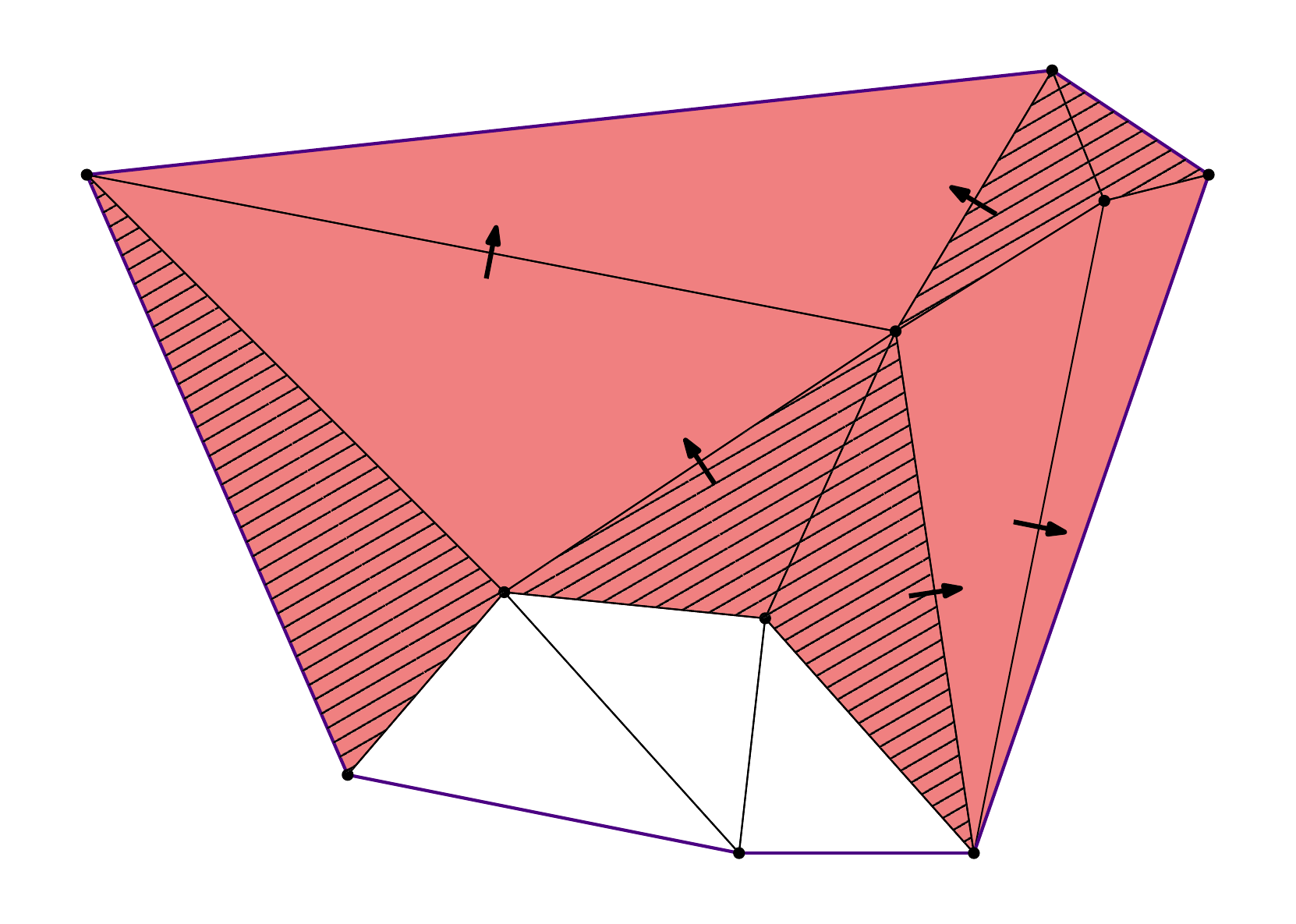}}}
\subfloat[\centering \nolinenumbers `\texttt{altitudeDrop}' insert action. \\$\mathrm{eval}(\mathcal{D})=17.8$, $\mathrm{eval}_3(\mathcal{D})=12.4$.\label{fig4:1}]{\scalebox{0.45}{\includegraphics[width=\textwidth]{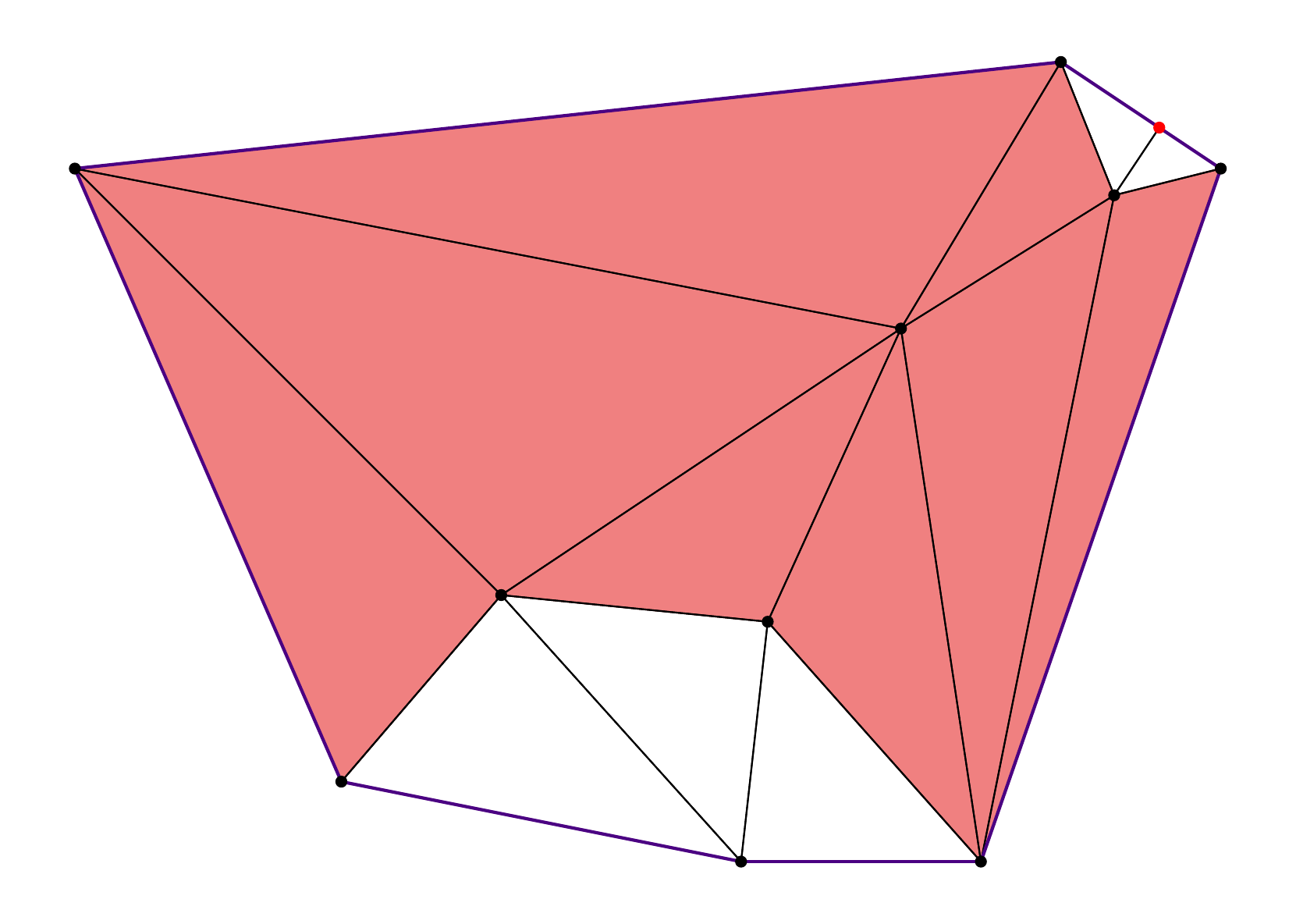}}}
    
    \subfloat[\nolinenumbers\centering `\texttt{altitudeDrop}' insert action. $\mathrm{eval}(\mathcal{D})=15.7$, $\mathrm{eval}_3(\mathcal{D})=11.2$.\label{fig4:2}]{\scalebox{0.45}{\includegraphics[width=\textwidth]{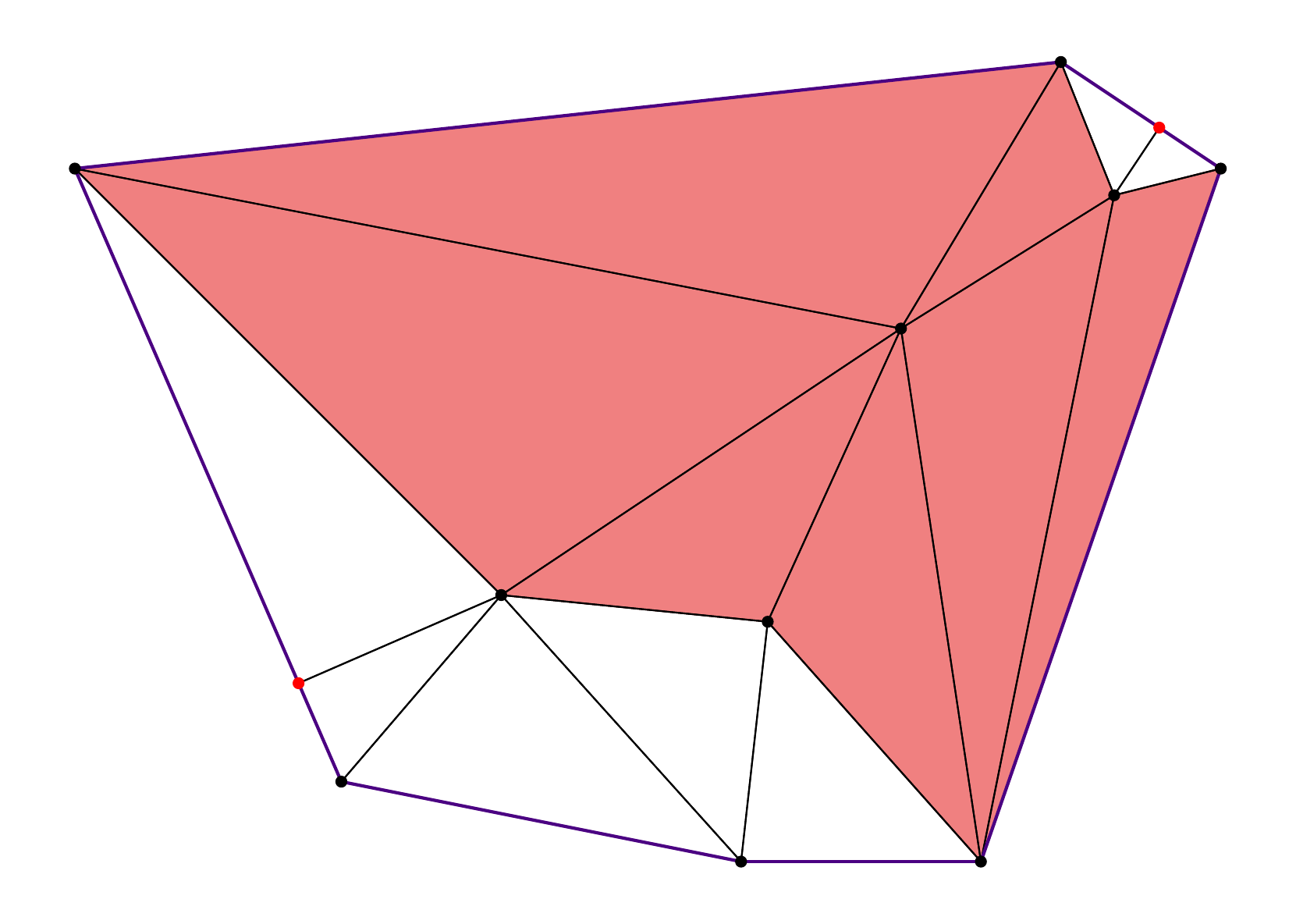}}}
    \subfloat[\centering \nolinenumbers `\texttt{visibilityVoronoi}' insert action. $\mathrm{eval}(\mathcal{D})=14.5$, $\mathrm{eval}_3(\mathcal{D})=9.1$.\label{fig4:3}]{\scalebox{0.45}{\includegraphics[width=\textwidth]{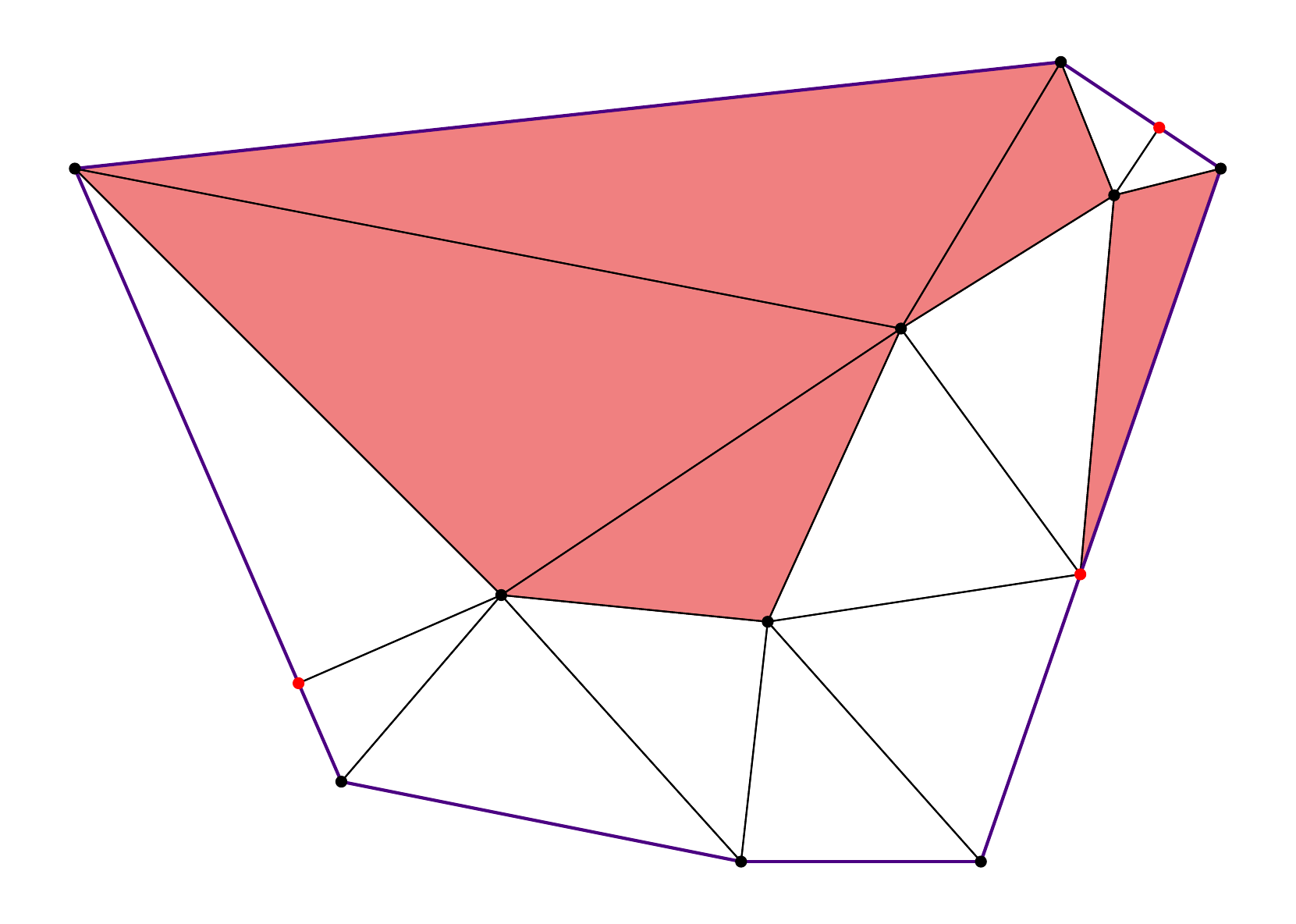}}}
    
    \subfloat[\nolinenumbers\centering `\texttt{altitudeDrop}' insert action. $\mathrm{eval}(\mathcal{D})=12.4$, $\mathrm{eval}_3(\mathcal{D})=7$.\label{fig4:4}]{\scalebox{0.45}{\includegraphics[width=\textwidth]{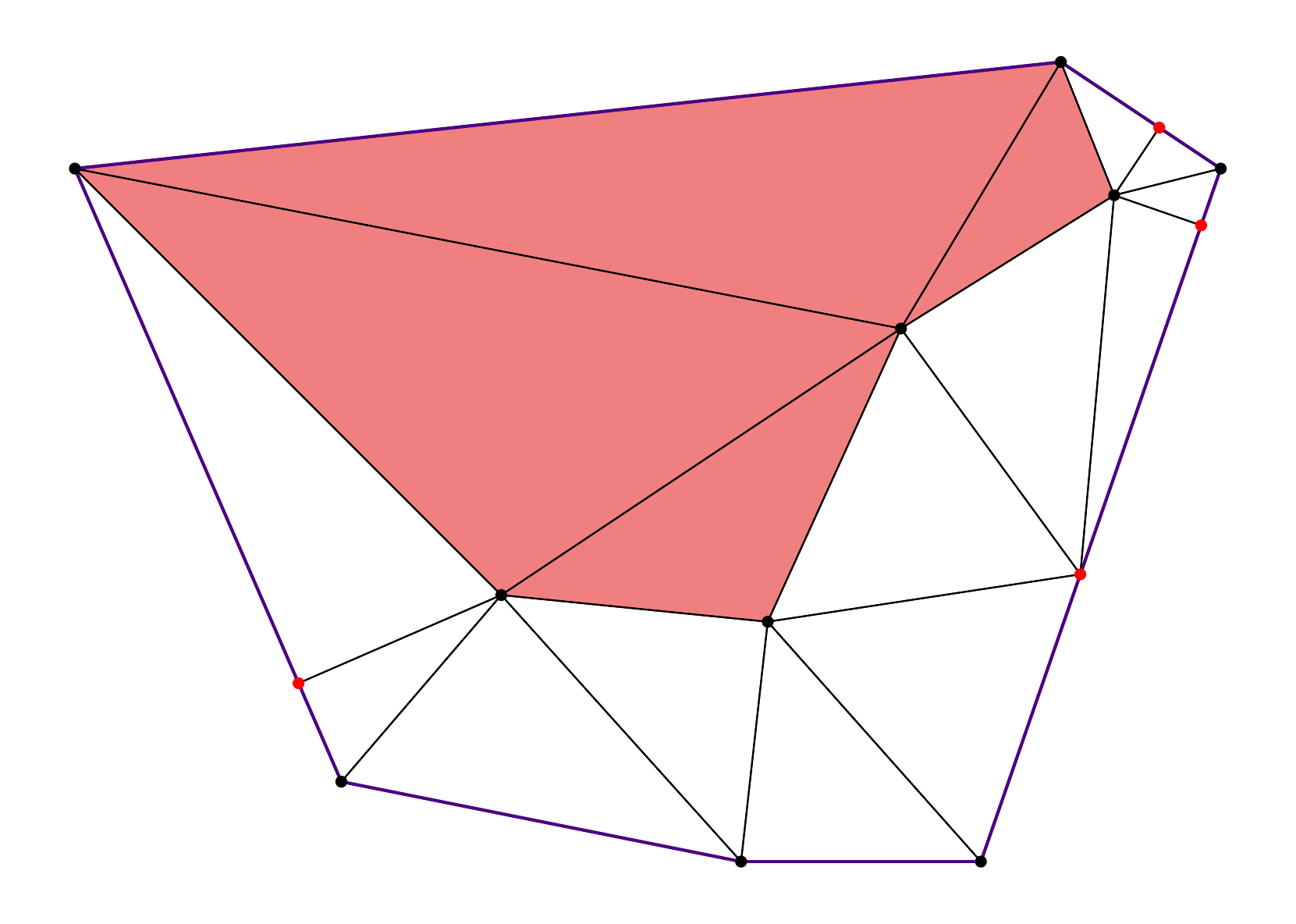}}}
    \subfloat[\centering \nolinenumbers `\texttt{polygonCenter}' insert action. $\mathrm{eval}(\mathcal{D})=11.2$, $\mathrm{eval}_3(\mathcal{D})=7$.\label{fig4:5}]{\scalebox{0.45}{\includegraphics[width=\textwidth]{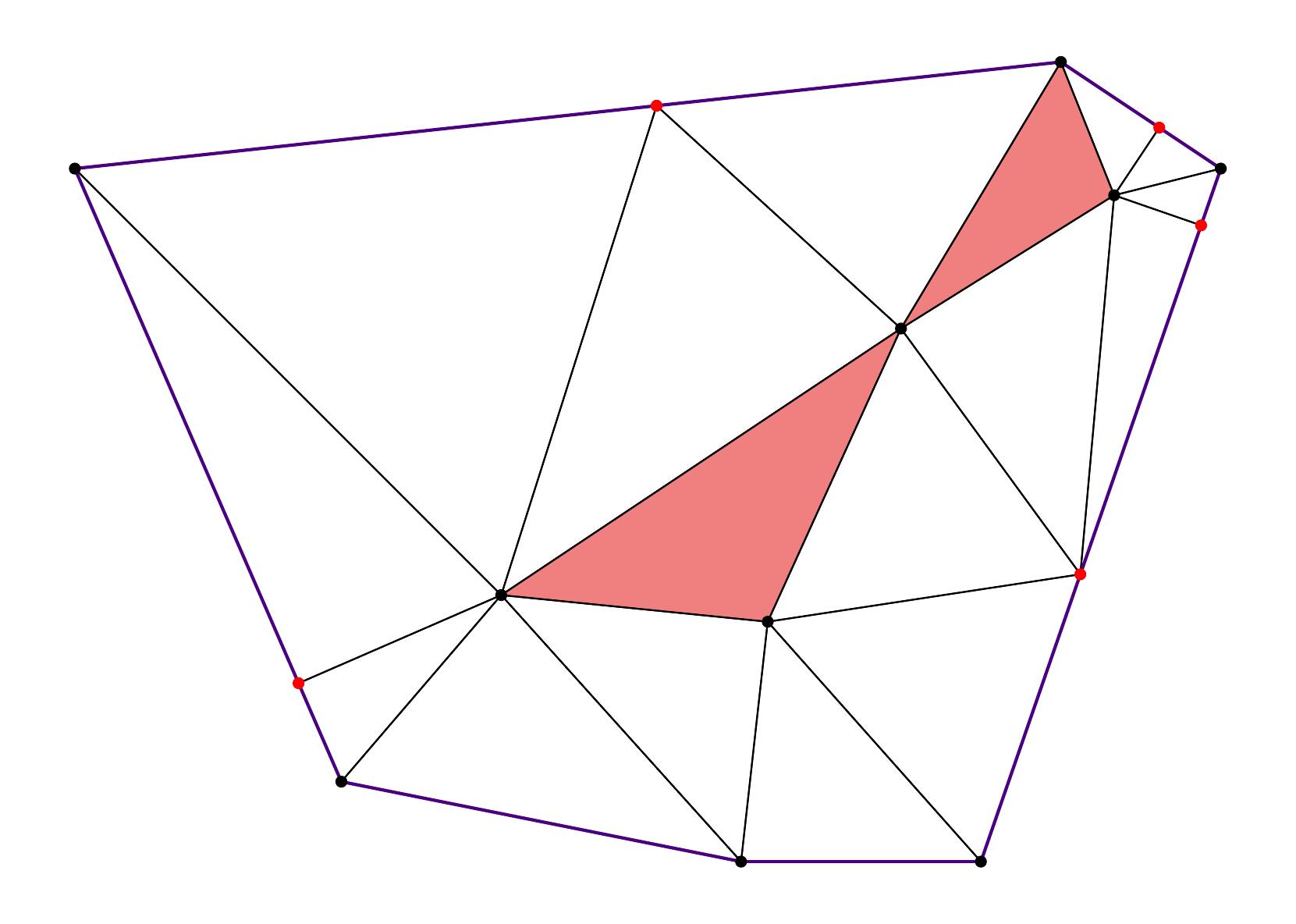}}}
    
    \subfloat[\nolinenumbers\centering `\texttt{polygonCenter}' insert action. $\mathrm{eval}(\mathcal{D})=9.1$, $\mathrm{eval}_3(\mathcal{D})=7$.]{\scalebox{0.45}{\includegraphics[width=\textwidth]{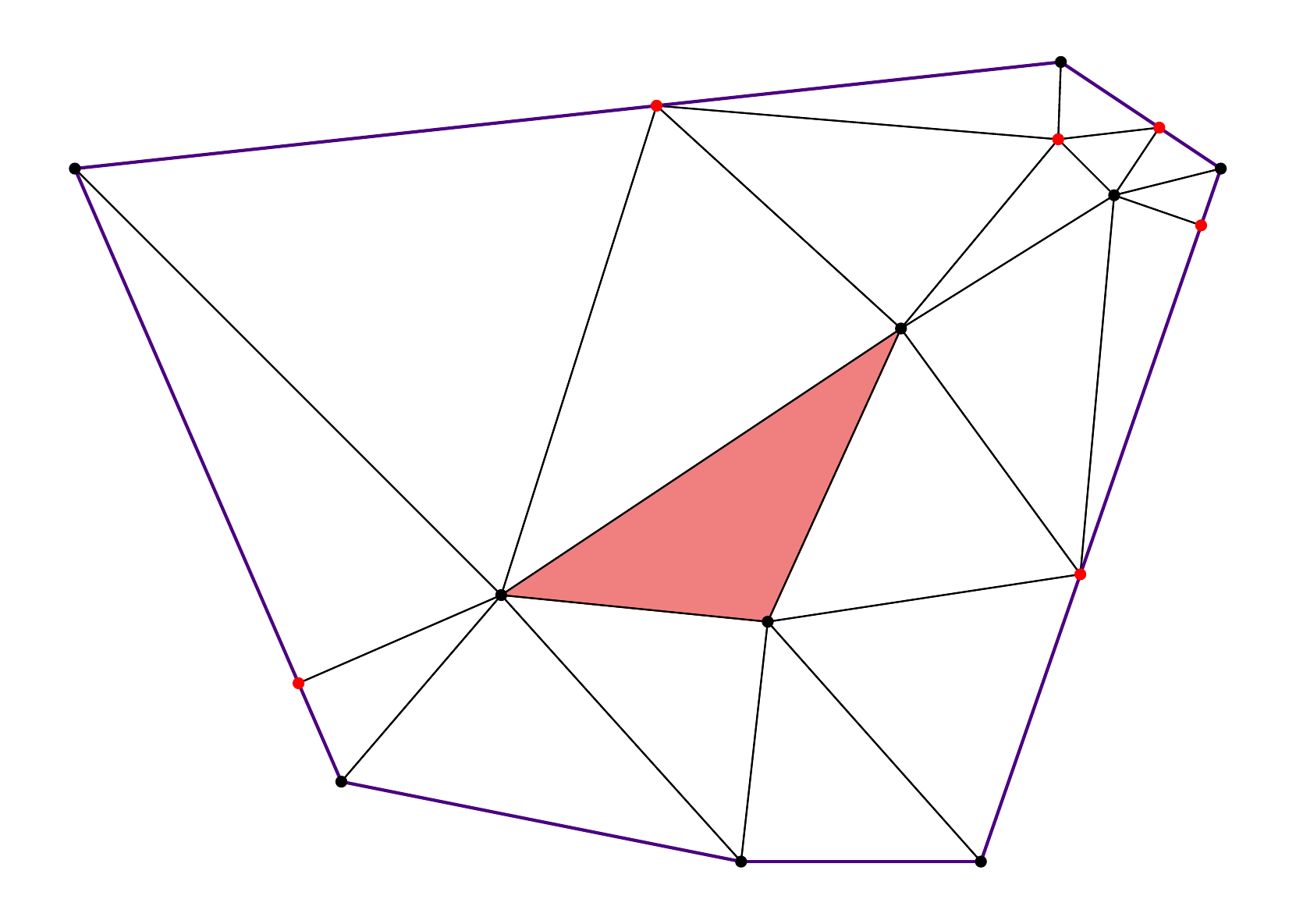}}}
    \subfloat[\centering \nolinenumbers Terminal `\texttt{polygonCenter}' insert action. $\mathrm{eval}(\mathcal{D})=7$, $\mathrm{eval}_3(\mathcal{D})=7$.\label{fig4:7}]{\scalebox{0.45}{\includegraphics[width=\textwidth]{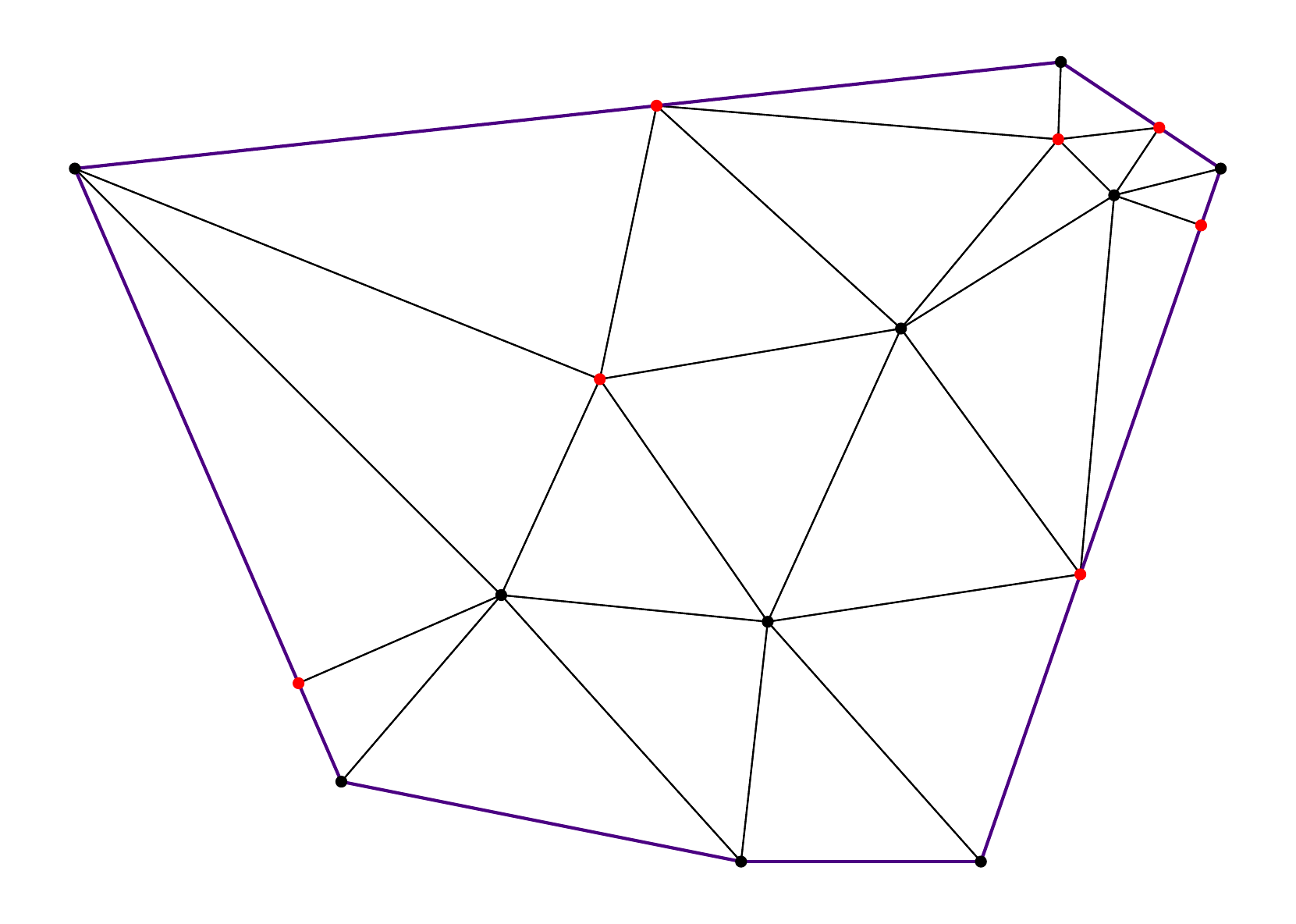}}}
    
    \caption{Maintained CDT of instance `\texttt{point-set\_10\_c04b0024}' (consisting of $10$ points, $6$ polygon edges and $0$ inner edges) throughout the execution of the local search with depth $0$, with all obtuse triangles marked in red. Execution time: $10$ seconds. Final number of Steiner points: $7$.}
    \label{fig4}
\end{figure}

\subparagraph{Na\"ive evaluation} We say an obtuse triangle $T$ in $\mathcal{D}$ \textit{supersedes} another obtuse triangle $T'$ in $\mathcal{D}$ if the longest side of $T$ is either of the shorter two sides of $T'$ (Figure \ref{fig4:0} shows superseded-superseding triangle pairs).
We expect superseded triangles to be influenced by whatever action is associated to its superseding triangle. An example of this phenomenon can be seen in \Cref{fig:center-identification}.
Hence, we argue that handling non-superseded obtuse triangles should be prioritized and therefore the cost of non-superseded obtuse triangles should be higher than that of superseded (obtuse) triangles.
The evaluation function we use is
\begin{align}\label{eq:evalfunction}
\mathrm{eval}(\mathcal{D}) \coloneqq \#\text{Steiner points} + 1.1\cdot\#\text{superseded }\triangle\text{'s} + 3.1\cdot\#\text{non-superseded }\triangle\text{'s}.
\end{align}
\cref{fig4} shows a run of the algorithm, including the actions that were applied and the value of the evaluation function. 
 Note that we did not pursue optimizing the constants in~\eqref{eq:evalfunction}.

\begin{figure}[!ht]
    \centering
    \subfloat[\centering \nolinenumbers Instance `\texttt{point-set\_150\_982c9ab3}' with $150$ points, $13$ polygon- and  $0$ inner edges.\label{fig1:instance}]{\scalebox{0.48}{\includegraphics[width=\textwidth]{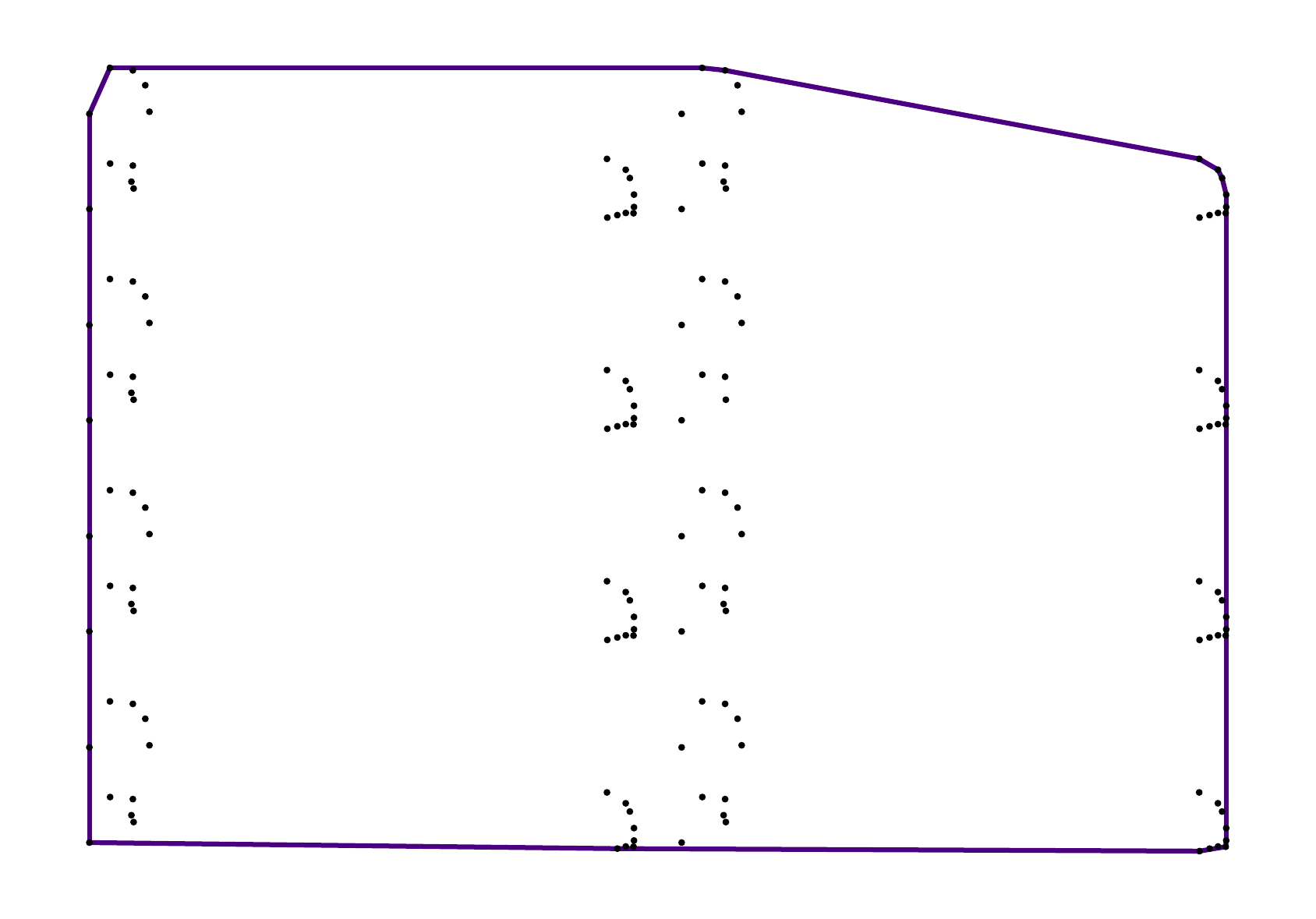}}}
    \subfloat[\centering \nolinenumbers Baseline solution via~\cite{ungörerten1,ungörerten2, erten-not} and rounding of coordinates with $573$ Steiner points.\label{fig1:baseline}]{\scalebox{0.48}{\includegraphics[width=\textwidth]{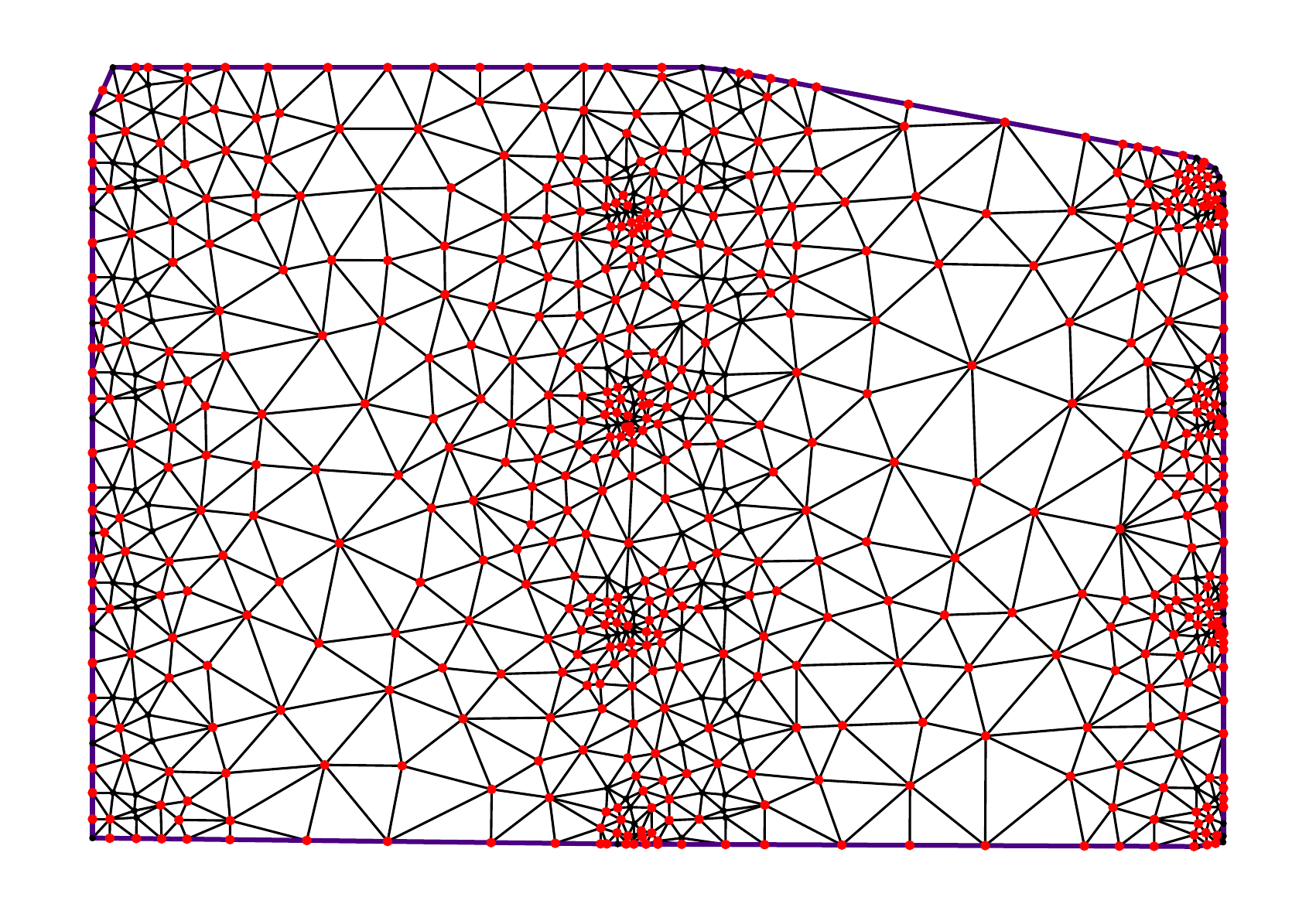}}}\\
    \subfloat[\centering \nolinenumbers Best found solution via probabilistic local search and depth $3$ with $190$ Steiner points.\label{fig1:localSearch}]{\scalebox{0.48}{\includegraphics[width=\textwidth]{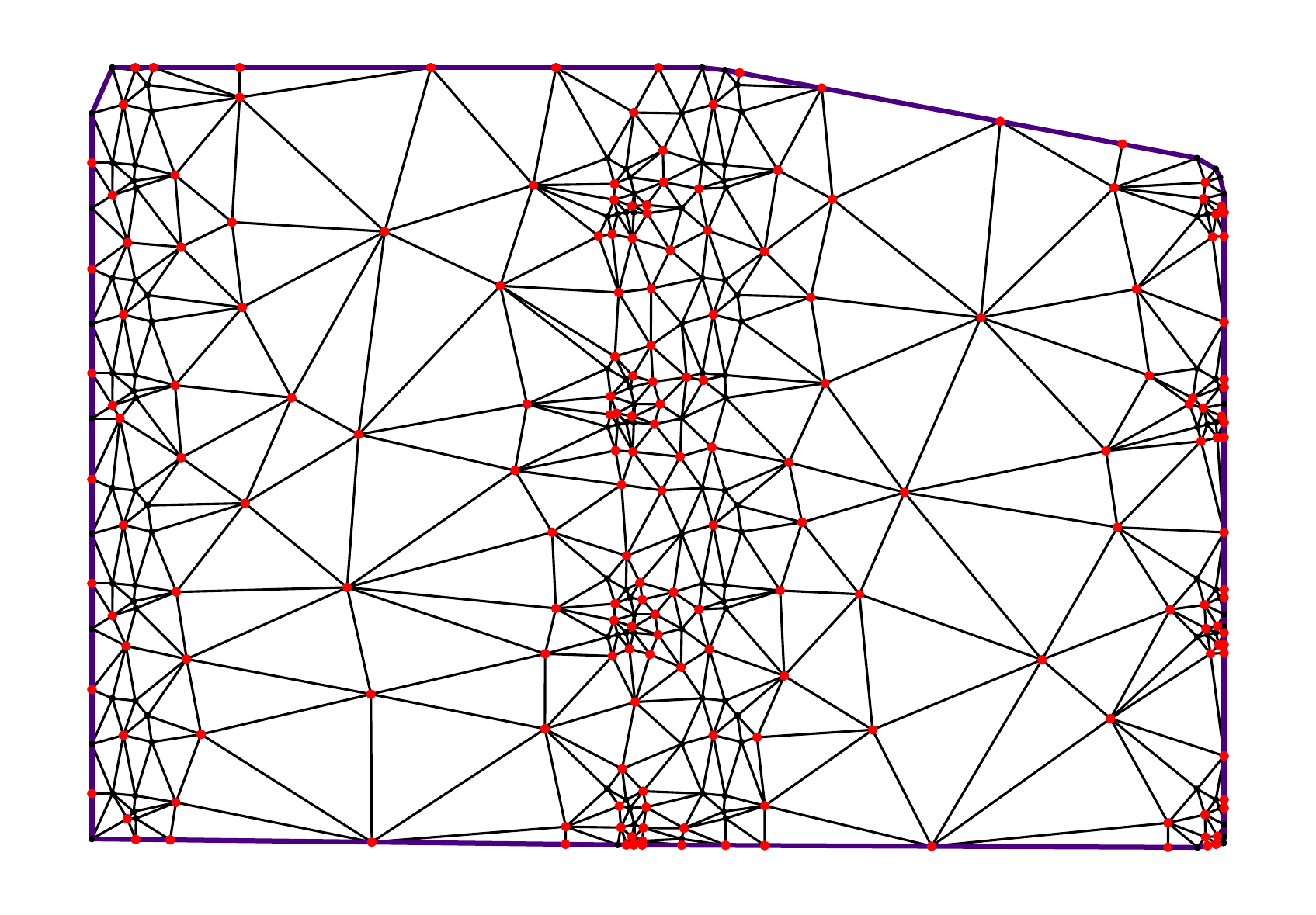}}}
    \subfloat[\centering \nolinenumbers Solution via merging and re-solving of existing solutions with $151$ Steiner points.\label{fig1:merge}]{\scalebox{0.48}{\includegraphics[width=\textwidth]{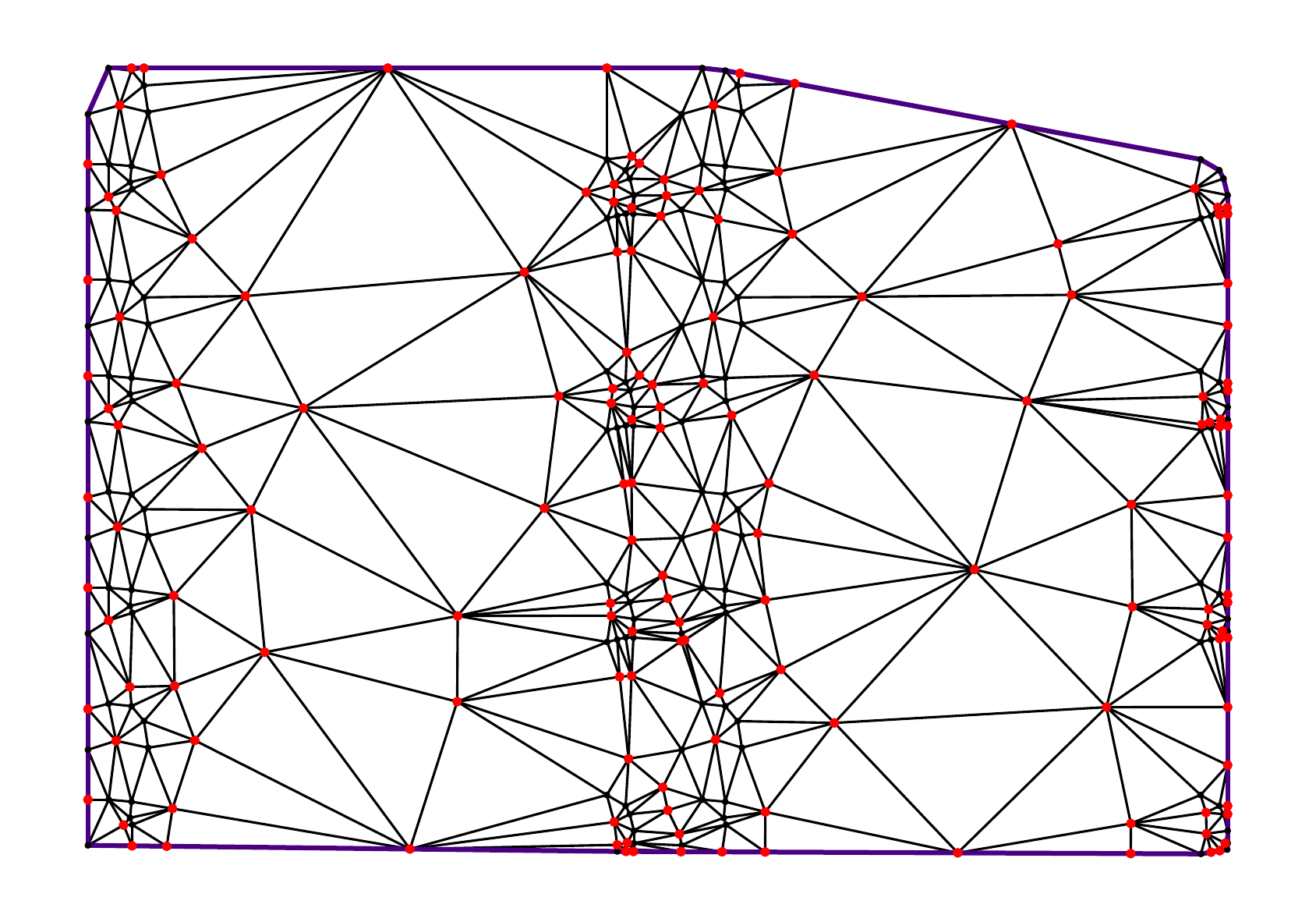}}}
    \caption{Solution comparison of instance `\texttt{point-set\_150\_982c9ab3}'.}
    \label{fig1}
\end{figure}

\subparagraph{Probabilistic state-graph evaluation} To improve the rather na\"ive evaluation function, we endow the evaluation function with a depth parameter $k\geq 0$, akin to the notion of minimax searching \cite{maschler_game_2013} over a game-state graph of a single player game, and define
\begin{equation}
  \mathrm{eval}_k(\mathcal{D}) \coloneqq 
    \begin{cases}
      \min_{A\in\mathcal{A}(\mathcal{D})} \mathrm{eval}_{k-1}(\mathcal{D}_A)& \text{if $k>0$}\\
      \mathrm{eval}(\mathcal{D}) & \text{if $k=0$}
    \end{cases}       
\end{equation}
where $\mathcal{D}_A$ is the CDT resulting from $\mathcal{D}$ when applying the action $A$ from the action set $\mathcal{A}(\mathcal{D})$. Even after extensive optimization efforts, we found that even with $k$ as small as $3$ the execution time was quite slow.
Hence, instead of evaluating $\mathrm{eval}_{k-1}(\mathcal{D}_A)$ for every $A$ in the action set of $\mathcal{D}$, we randomly draw a sample $S$ of actions from $\mathcal{A}(\mathcal{D})$ (with a bias towards more promising moves determined by $\mathrm{eval}_0(\mathcal{D}_A)$) computing $\min_{A\in S} \mathrm{eval}_{k-1}(\mathcal{D}_A)$.

\subsection{Solution Merging}
We observed that, due to the probabilistic nature of the evaluation function, different solutions to the same instance often had a similar number of Steiner points but looked vastly different locally. This motivates merging different solutions by sampling a circle and replacing every Steiner point inside this circle from one solution with the Steiner points of the other solution.
Note that merging two solutions might result in a CDT that has obtuse triangles.
Thus, we apply our algorithm to these CDTs obtaining a new solution with potentially less Steiner points. Repeated application of this merging step reduced the number of Steiner points by as much as $30\%$ for some instances (see \Cref{fig1}).

\section{Implementation, Experiments, and Examples}\label{sec:implementation}

\subparagraph*{Code and Hardware}
Our implementation is in python.
In the later stages of the competition we ran our solver on ten nodes of the Marvin cluster of the University of Bonn, each with $1024\mathrm{GB}$ of RAM and $96$ threads at $2.10\mathrm{GHz}$ using around $300\ 000$ CPU hours.
We did not find any improved solution in the last two weeks of the competition.

\begin{figure}[!t]
    \centering
    \includegraphics[width=\textwidth]{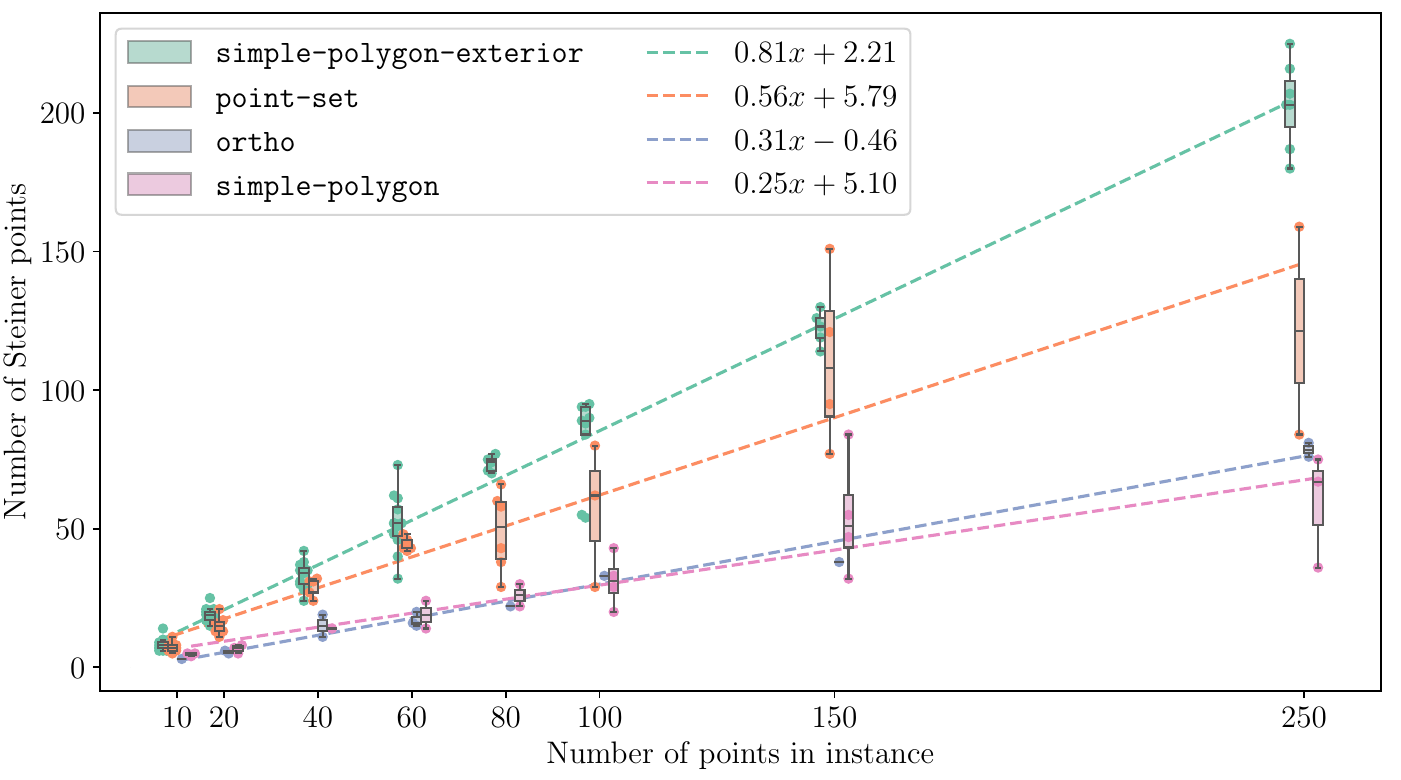}
    \caption{Relationship between solution size and input complexity over different instance types.}
    \label{fig2}
\end{figure}

\begin{figure}
    \centering
    \includegraphics[width=\textwidth]{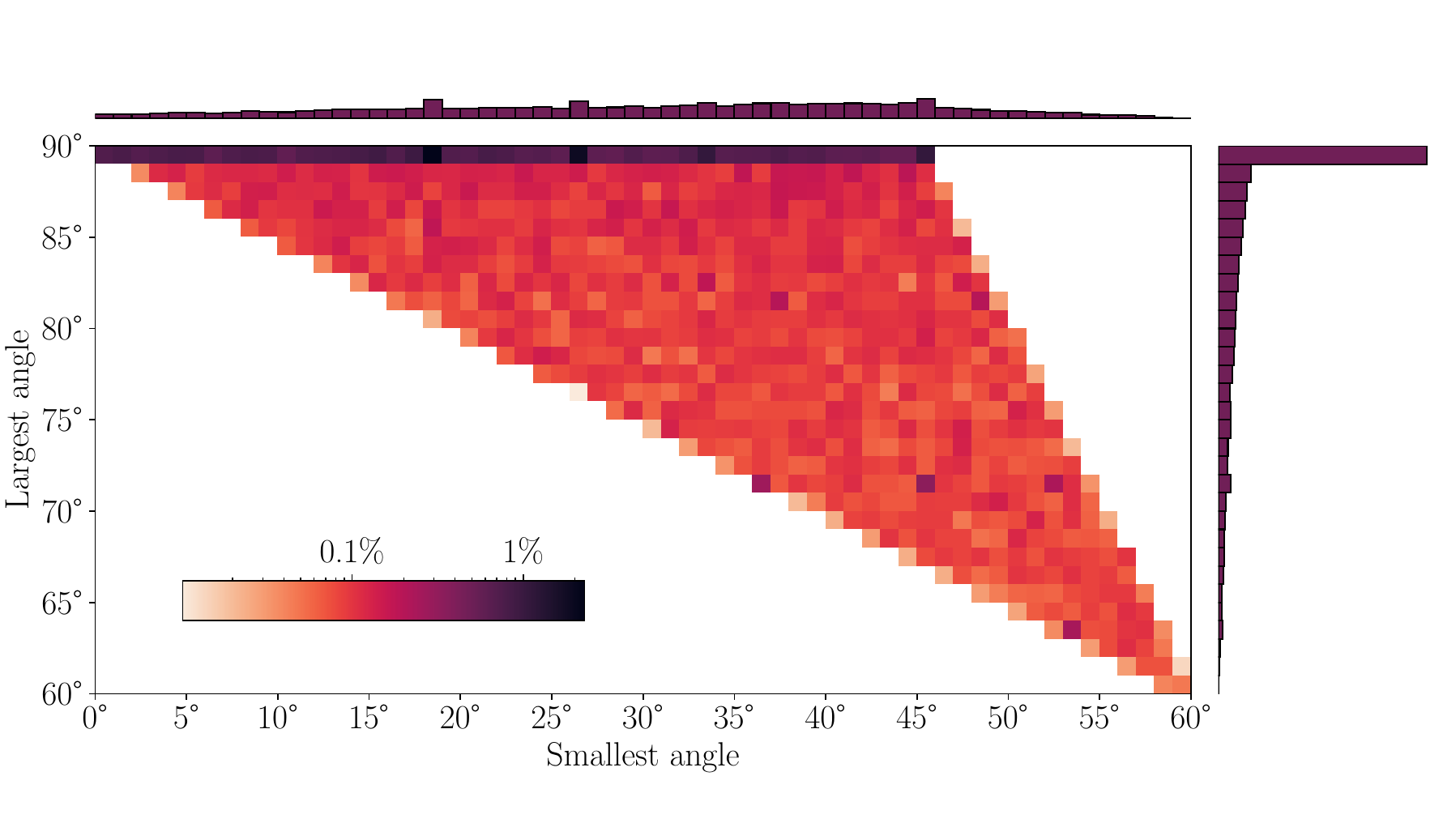}
    \caption{Average distribution of smallest and largest angles, of all triangles in all solutions.}\label{fig:angle_distribution}

\end{figure}

\begin{figure}
    \centering
    \includegraphics[width=0.48\linewidth]{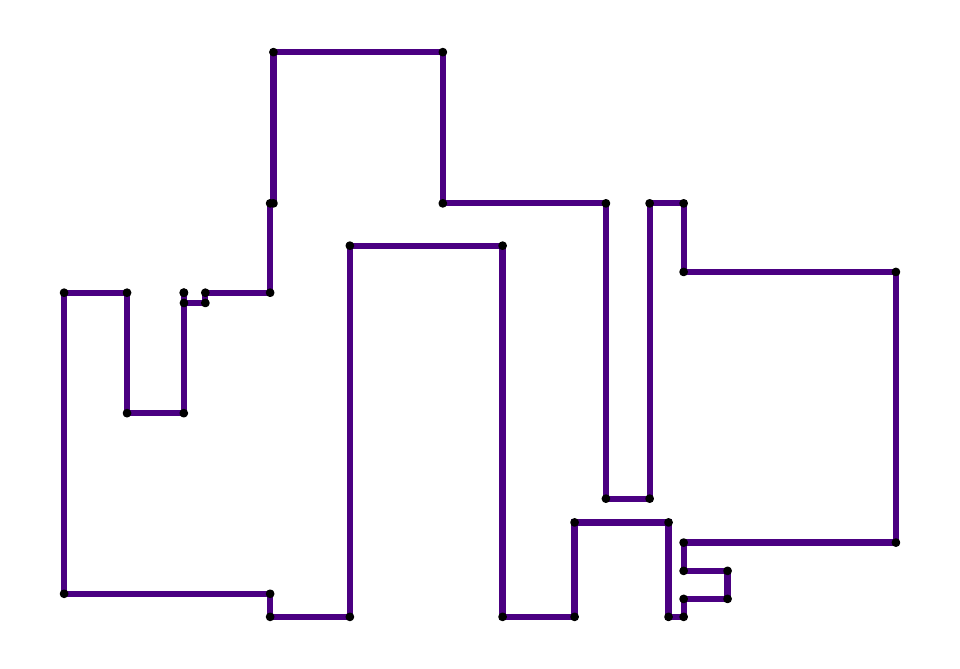}
    \hfill
    \includegraphics[width=0.48\linewidth]{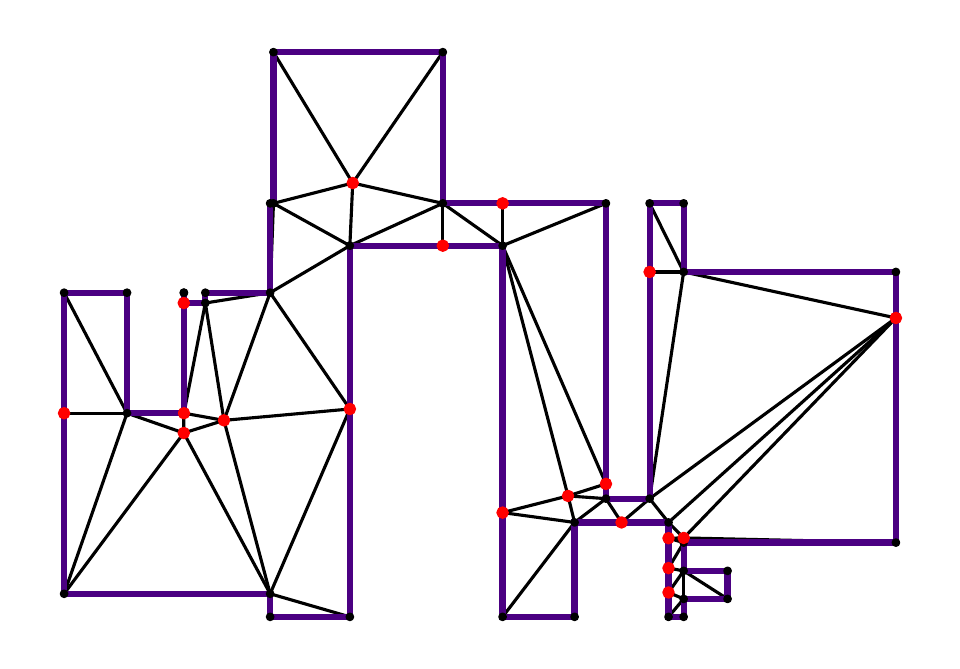}
    \caption{Instance `\texttt{ortho\_40\_56a6f463}' and a solution with $19$ Steiner points.}
    \label{fig:ortho-instance}
\end{figure}

\begin{figure}
    \centering
    \includegraphics[width=0.45\linewidth]{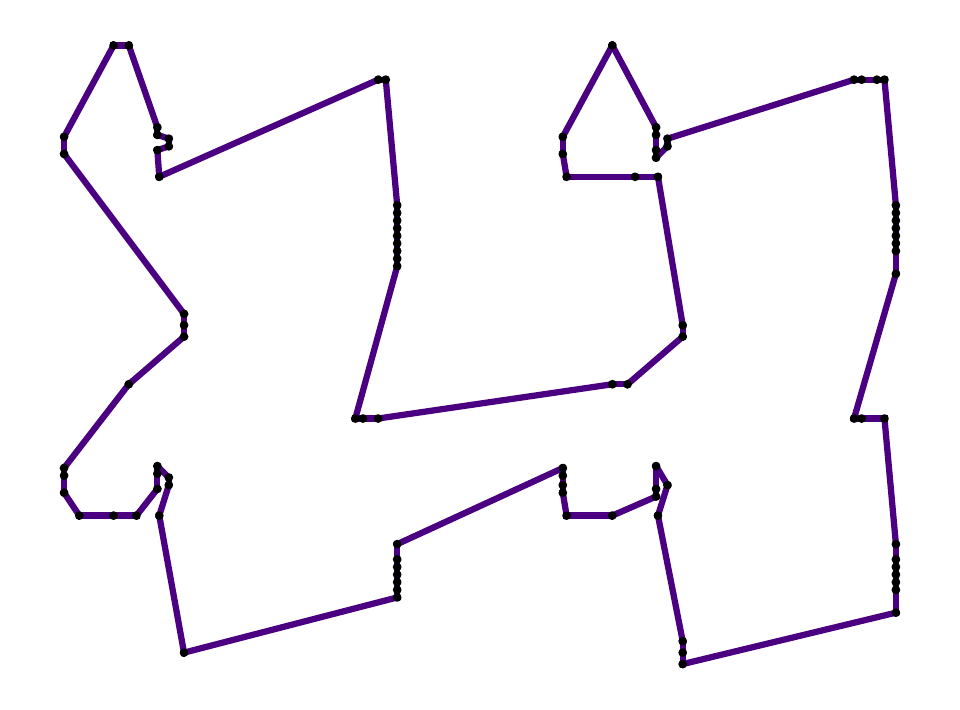}
    \hfill
    \includegraphics[width=0.45\linewidth]{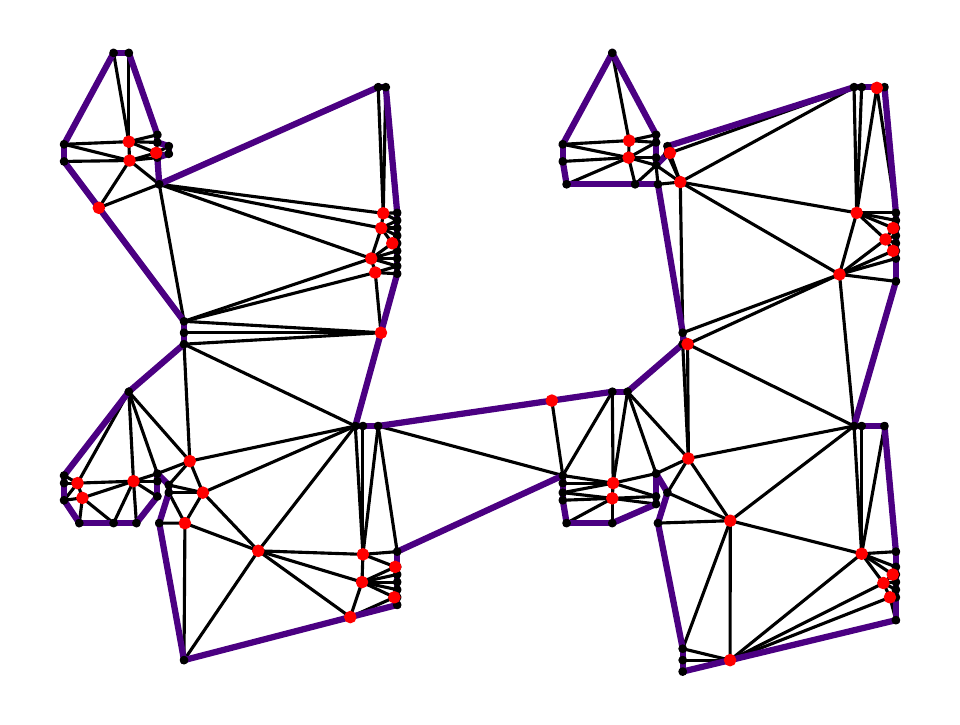}
    \caption{Instance `\texttt{simple-polygon\_100\_cb23308c}' and a solution with $43$ Steiner points.}
    \label{fig:simple-polygon-instance}
\end{figure}

\begin{figure}
    \centering
    \includegraphics[width=0.45\linewidth]{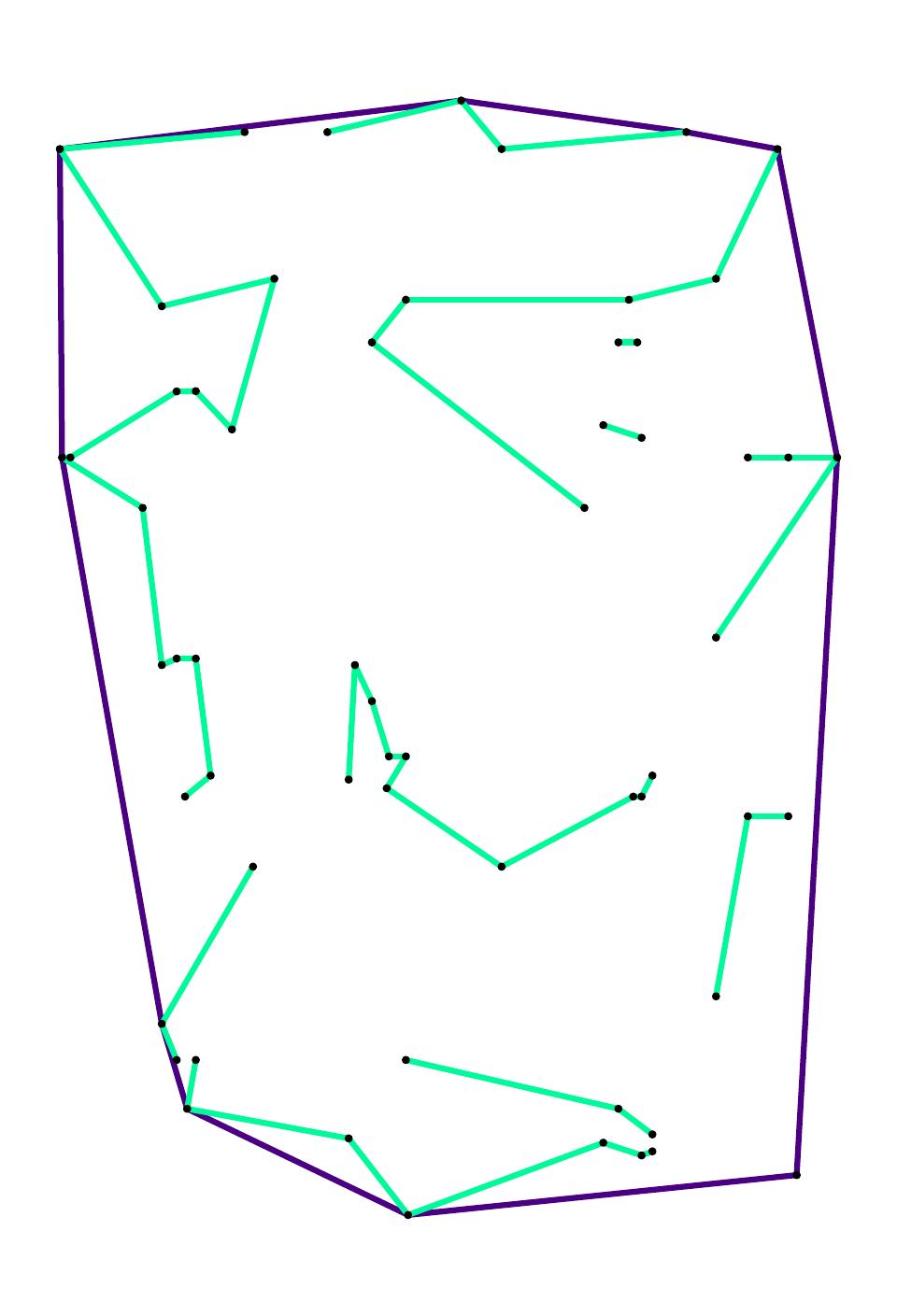}
    \includegraphics[width=0.45\linewidth]{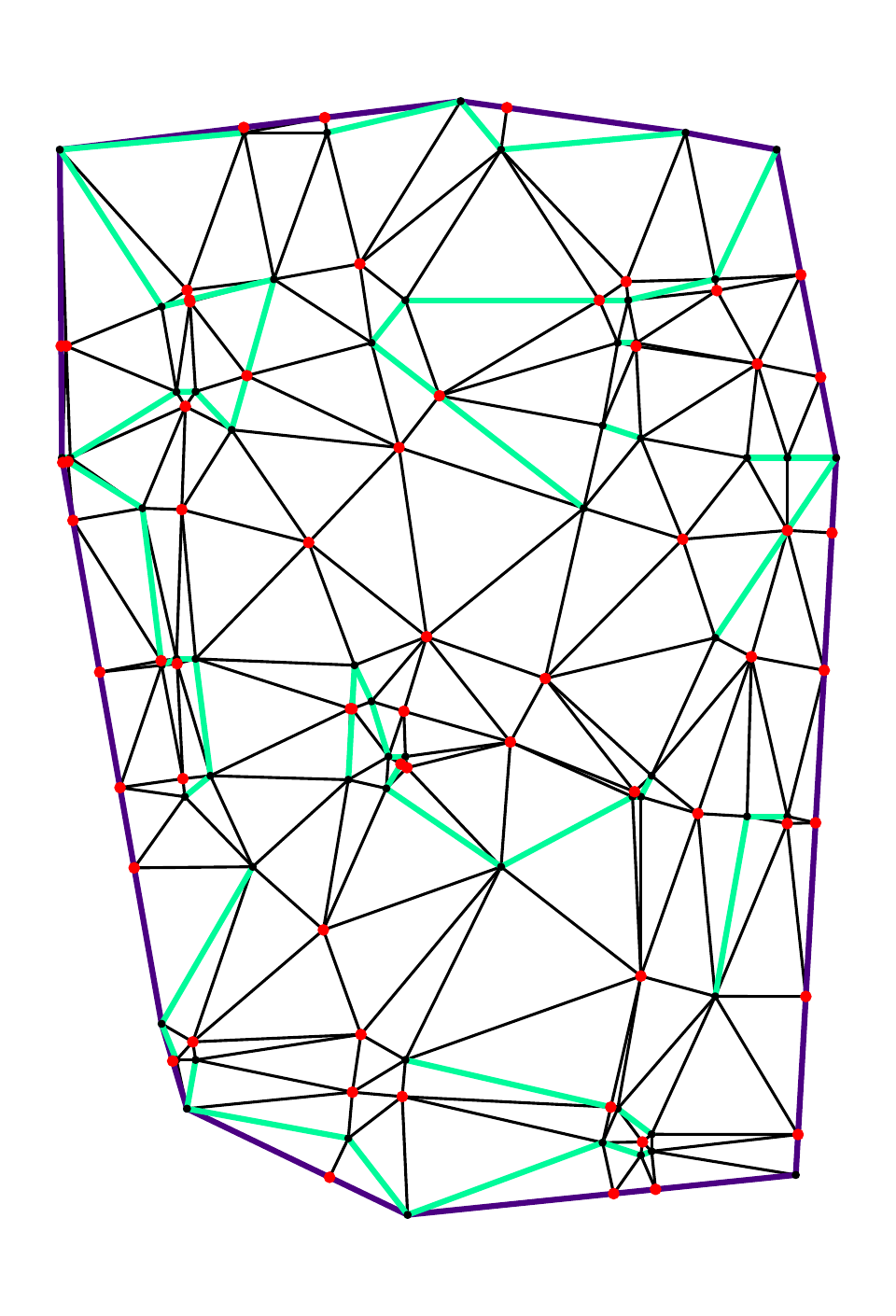}
    \caption{Solution to `\texttt{simple-polygon-exterior-20\_60\_8221c868}' with $62$ Steiner points.}
    \label{fig:polygon-exterior-2}
\end{figure}

\begin{figure}
    \centering
    \includegraphics[width=0.48\linewidth]{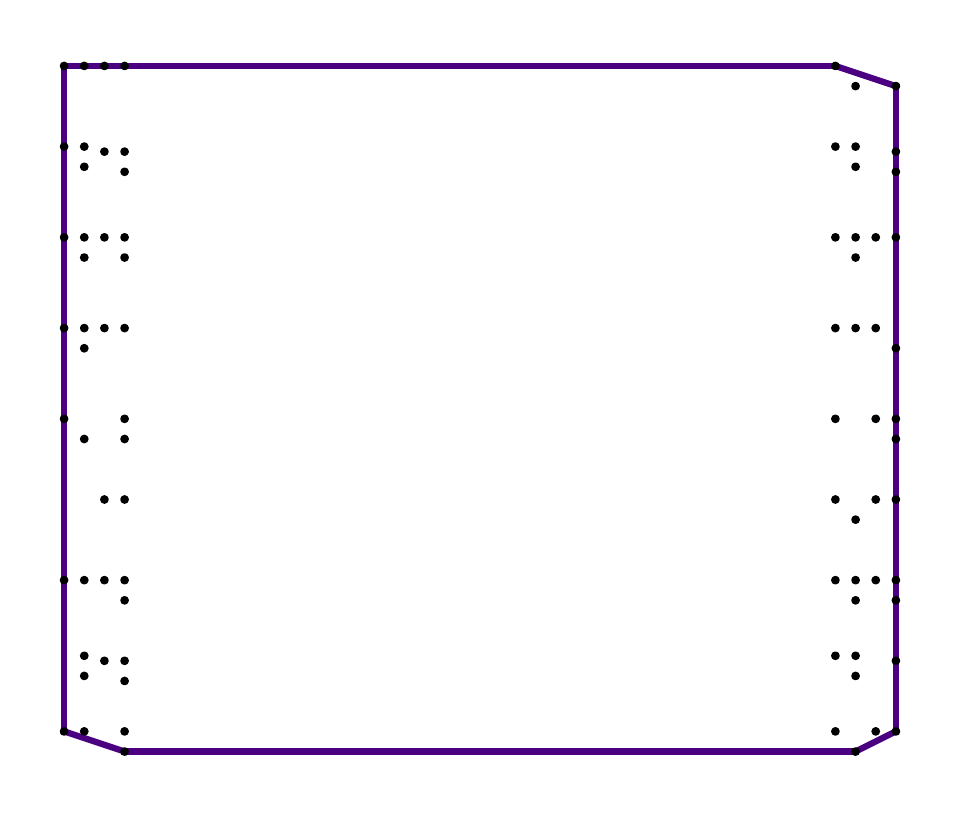}
    \hfill
    \includegraphics[width=0.48\linewidth]{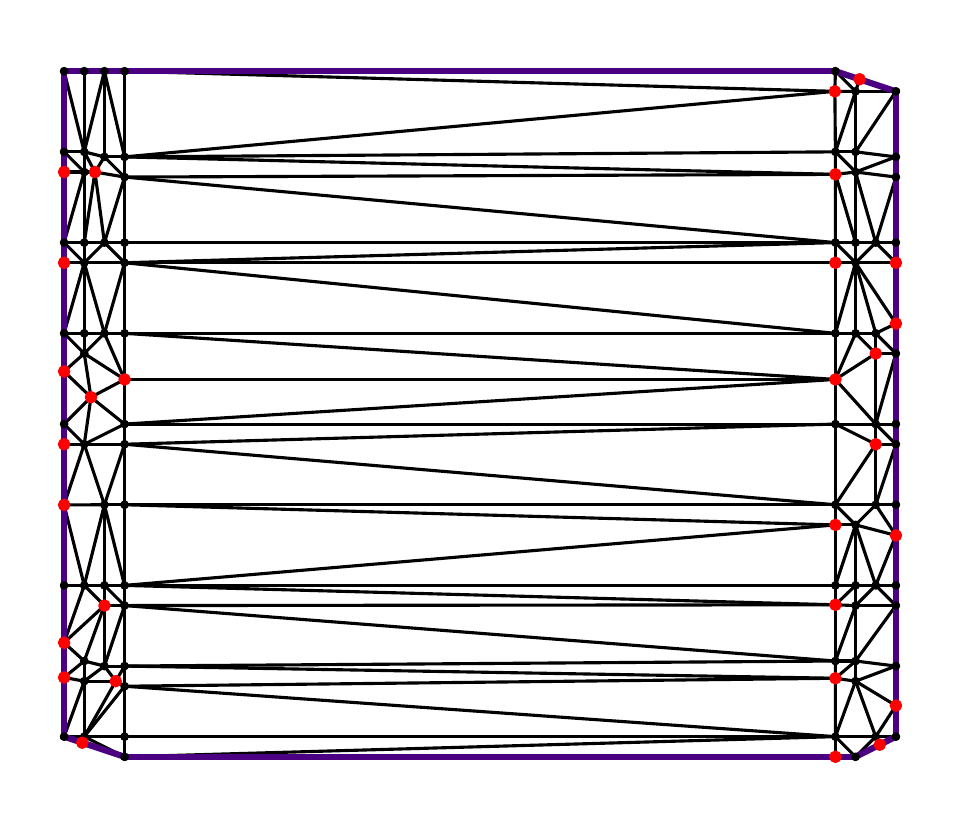}
    \caption{Instance `\texttt{point-set\_80\_9a8373fb}' and a solution with $29$ Steiner points.}
    \label{fig:point-set-1}
\end{figure}

\begin{figure}
    \centering
    \includegraphics[width=0.48\linewidth]{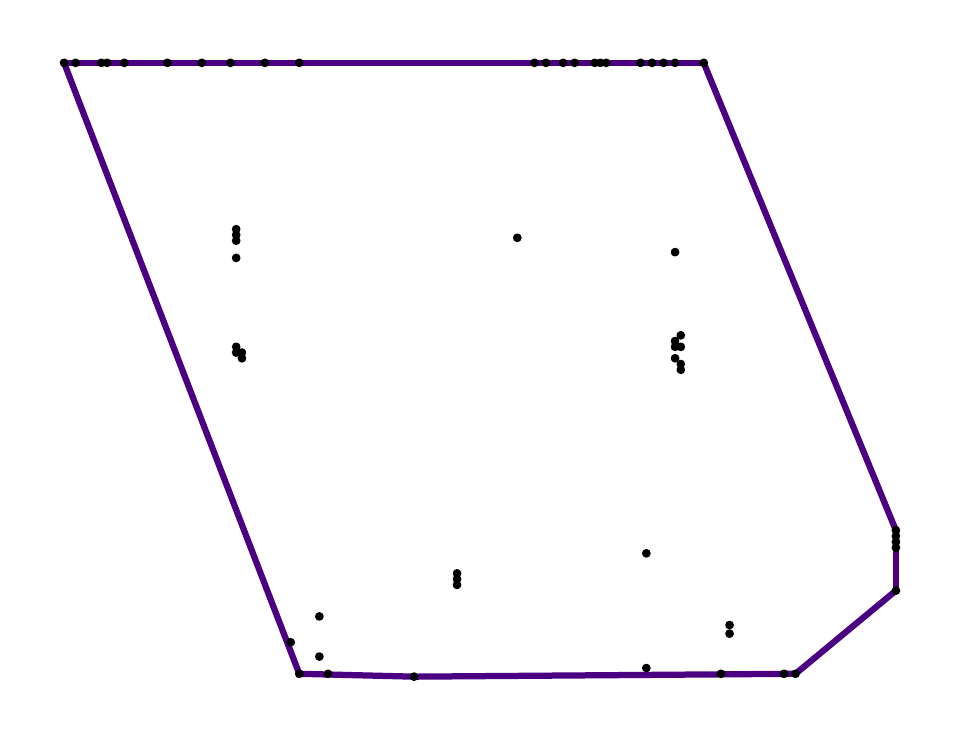}
    \hfill
    \includegraphics[width=0.48\linewidth]{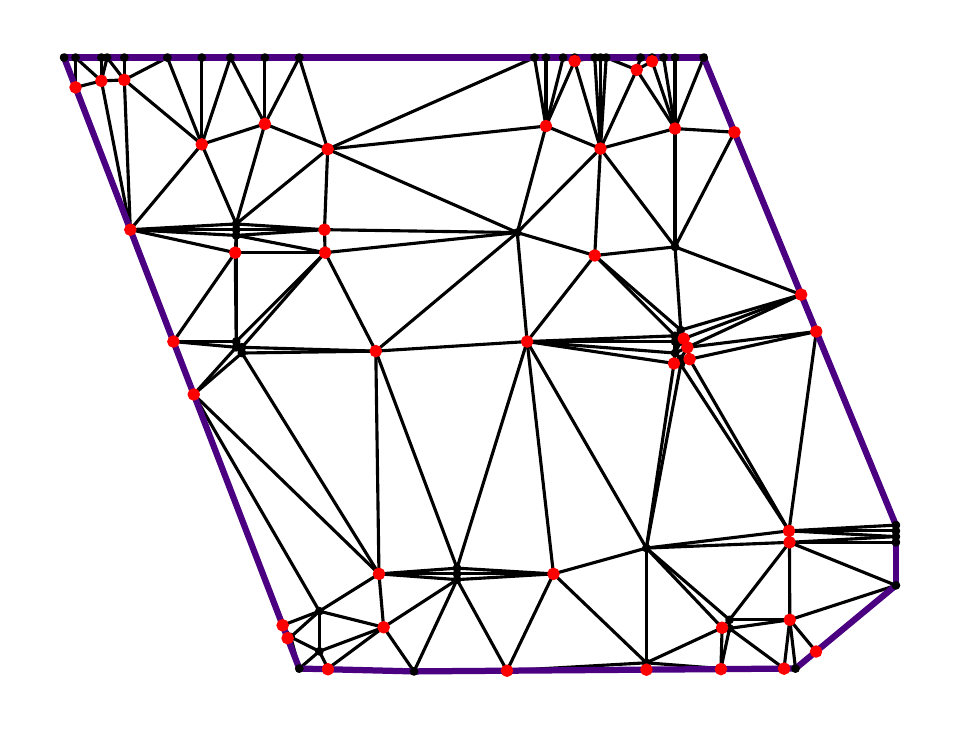}
    \caption{Instance `\texttt{point-set\_60\_27bc003d}' and a solution with $43$ Steiner points.}
    \label{fig:point-set-2}
\end{figure}

\begin{figure}
    \centering
    \includegraphics[width=\linewidth]{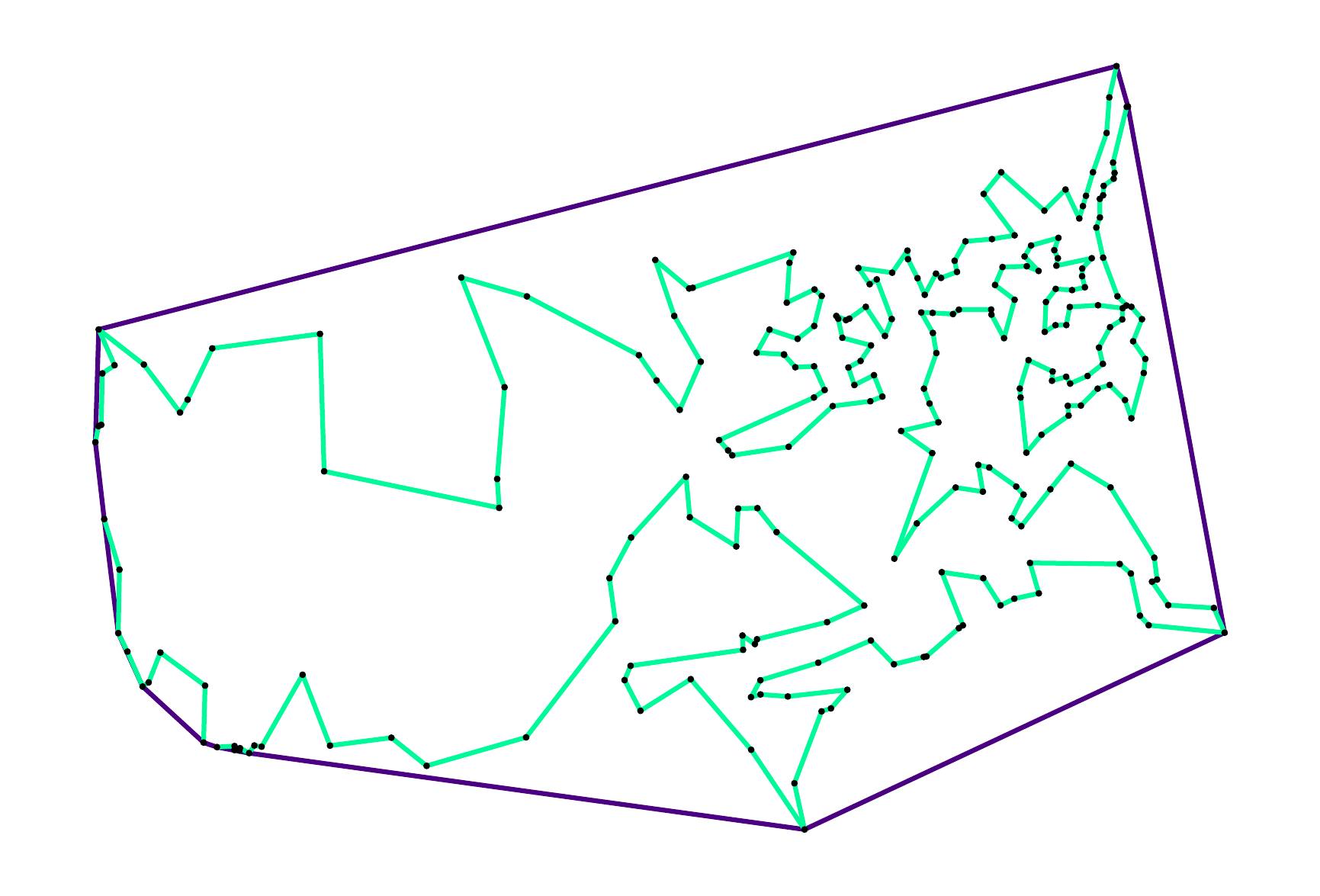}\\
    \includegraphics[width=\linewidth]{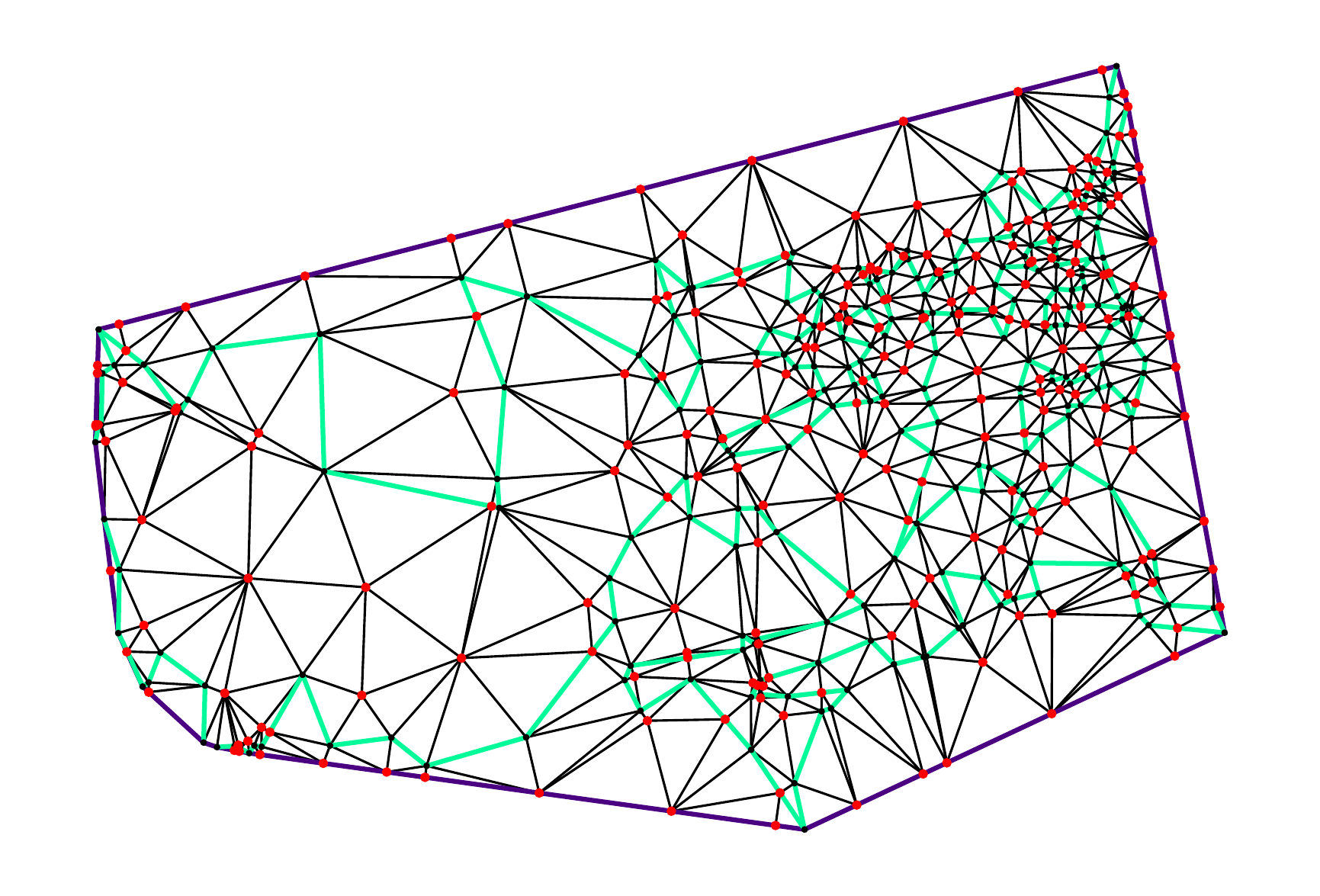}
    \caption{Solution with $225$ Steiner points to `\texttt{simple-polygon-exterior\_250\_c0a19392}'.}
    \label{fig:polygon-exterior}
\end{figure}

\subparagraph*{Solutions}
The provided instances fall into various types: (i) \texttt{ortho}, consisting of rectilinear simple polygons (\Cref{fig:ortho-instance}), (ii) \texttt{simple-polygon}, consisting of simple polygons (\Cref{fig:simple-polygon-instance}), (iii) \texttt{point-set}, consisting of the convex polygons with points in its interior (\Cref{fig:point-set-1,fig:point-set-2}), and (iv) \texttt{simple-polygon-exterior}, consisting of convex polygons with points and additional constraints in its interior (\Cref{fig:polygon-exterior,fig:polygon-exterior-2}).
Our algorithm performs well on all instances types; no manual intervention or distinction is necessary for our approach.
The number of Steiner points in our solutions compared to the number of input points in each type suggests that the total number of Steiner points is roughly linear in the input complexity and increases depending on the number of points and constraints in the interior of the PSLG, see \Cref{fig2}.
We observe that a large fraction of triangles computed by our algorithm are right triangles (see \Cref{fig:angle_distribution}, right).
All other angles in our solutions (determined by largest and smallest angle) appear to be roughly equally distributed (see \Cref{fig:angle_distribution}).
\newpage
\newpage

\clearpage

\bibliography{bibliography.bib}

\end{document}